\documentclass{amsart}
\usepackage{graphicx}
\usepackage{color}
\usepackage{tikz}
\usepackage{graphicx}
\usepackage{graphicx}
\usepackage{amssymb}
\usepackage{amsfonts}
\usepackage{amsmath}
\usepackage{amsthm}
\newtheorem{Theorem}{Theorem}[part]
\newtheorem{Definition}{Definition}[part]

\newtheorem{Lemma}{Lemma}[part]
\newtheorem{Corollary}{Corollary}[part]
\newtheorem{Remark}{Remark}[part]
\newtheorem{Example}{Example}[part]

\newcommand{\be}{\begin{equation}}
\newcommand{\ee}{\end{equation}}

\numberwithin{equation}{section}

\newcommand{\e}{{\epsilon}}
\begin{document}

\title{Additive habits with power utility: Estimates, asymptotics and equilibrium}

 \author{Roman Muraviev\\
 Department of Mathematics and RiskLab\\
 ETH Zurich\\
 }%

\address{Department of Mathematics and RiskLab, ETH Zurich,
Zurich 8092, Switzerland.
\newline
 {e.mail: roman.muraviev@math.ethz.ch}}

 \date{\today}
\begin{abstract}
We consider a power utility maximization problem with additive habits in a framework of discrete-time markets and random endowments. 
For certain classes of incomplete markets, we establish estimates for the optimal consumption stream in terms of the aggregate state price density, 
investigate the asymptotic behavior of the propensity to consume (ratio of the consumption to the wealth), as the initial endowment tends to infinity, 
and show that the limit is the corresponding quantity in an artificial market.
For complete markets, we concentrate on proving the existence of an Arrow-Debreu equilibrium in an economy inhabited by heterogeneous 
individuals who differ with respect to their risk-aversion coefficient, impatience rate and endowments stream, but possess the same degree of 
habit-formation. Finally, in a representative agent equilibrium, we compute explicitly the price of a zero coupon bond and the Lucas tree equity, 
and study its dependence on the habit-formation parameter.
\end{abstract}

\maketitle \markboth{R. Muraviev}{Additive Habits: Estimates, Asymptotics and Equilibrium.
}
\renewcommand{\theequation}{\arabic{section}.\arabic{equation}}
\pagenumbering{arabic}

\section{Introduction}

The classical problem of an investor optimizing his preference functional by selecting a suitable consumption plan
constitutes a significant topic in financial economics and mathematical finance. Since its origins dating back to the
seminal work of Merton \cite{MERTON}, the problem has attracted the attention of numerous researches (see e.g. \cite{DFSZ,KLSX,KZ,KS,MT,RS}),
causing a prominent progress in the development of novel mathematical tools, and an establishment of complex models which in particular aim to
appropriately explain important empirical observations.

One such modeling issue, which is a central ingredient in the current manuscript, is the habit-formation utility paradigm.
In contrast to standard time-separable utility functions,
habit preferences enjoy certain properties which are beneficial from an economic and psychological viewpoint. Namely,
in this model, the past consumption patterns of an individual carry an impact on his current policy. The intuition behind this model
is based on the postulation that decision
makers who consume portions of their wealth over time are supposed to develop habits, which will have a firm impact on their subsequent
consumption behavior. In particular, the relative desire to consume may be increased if one has become accustomed to high levels of consumption.
A vast range of works are devoted to the study of various aspects of the habit-forming utility maximization problem
(see \cite{A,CH,CHAPMAN,CO,DK,DZ1,DZ2,EK,MUR}).

The present manuscript deals with an individual's discrete-time power utility optimization
problem with additive habits. At each period, the current consumption choice is subtracted
from a benchmark parameter, which is commonly referred to in the literature as the standard of living index,
and is equal to a weighted average of the past consumption stream. Due to the fact that power utility functions are defined
on the set of non-negative real numbers, the individual is forced to consume in an addictive manner, since he is not permitted
to consume below the benchmark level.

The article can be categorized into two parts, which can be read independently. In the first part (Sections 3 and 4), various classes of 
incomplete markets are considered: arbitrary incomplete
markets with a deterministic interest rate, idiosyncratically incomplete markets (see Definition \ref{def1}) and markets of type $\mathcal{C}$
(introduced in Malamud and Trubowitz \cite{MT}, see Definition \ref{marketc}). By exploiting the characterization of the solution of the 
habit-forming maximization problem in the setting of the preceding markets, which was developed in Muraviev \cite{MUR}, we provide estimates 
for the optimal consumption stream in terms of the aggregate state price density. 
Furthermore, we investigate the asymptotic behavior of the ratio of the optimal consumption policy to the wealth (propensity to consume), 
as the initial endowment tends to infinity, show that the corresponding limit is equal to the propensity to consume in an artificial market, 
and derive the convergence rate in the setting of some concrete markets. The second part (Section 5) is concerned with a complete market Arrow-Debreu 
equilibrium. We first provide explicit formulas for the so-called representative agent models, that is, a homogeneous economy. 
We then derive the price of a zero coupon bond and the Lucas tree equity, and prove that these prices are increasing convex
functions of the habit-formation coefficient.  
Secondly, we analyze and prove the existence of an equilibrium for a finite set of heterogeneous individuals that have distinct
risk-aversion coefficients, impatience rates and endowments, but coincide in the degree of their habits. The reader is addressed to
\cite{D1,D2,DZ1,KLS,M2} for an equilibrium related literature.

The paper is organized as follows. In section 2, we introduce all the essential notions and the introductory results. Section 3 is devoted
to the derivation of estimates for the optimal consumption stream. In section 4, we investigate the asymptotic behavior of the optimal consumption.
Section 5 concludes the paper with the analysis of an Arrow-Debreu equilibrium.

\section{Setup and Preliminaries}
The setup coincides with the one in Muraviev \cite{MUR}. We briefly depict the main concepts of the model. There are $T+1$ periods.
Uncertainty is characterized by a finite probability space $\left( \Omega, \mathcal{G}, P \right)$ and a filtration
$ \mathcal{G}_0 := \{ \emptyset, \Omega \}
\subseteq \mathcal{G}_1 \subseteq .... \subseteq \mathcal{G}_T := G$. We set $L^{2} \left(
\mathcal{G}_k \right)$, $k=0,...,T,$ to be the finite-dimensional space of all $\mathcal{G}_k$ measurable random variables,
endowed with the inner product
$ \langle X , Y \rangle := E \left[ X Y\right] $, for $X, Y \in L^{2} \left( \mathcal{G}_k \right). $
We set further $R_{+}:=[0, \infty),$ $R_{++}:=(0, \infty),$ $L^{2}_{+} \left( \mathcal{G}_k \right) := \left\{ X \in L^{2} \left( \mathcal{G}_k \right)
\big| X \geq 0 \right\}$ and $L^{2}_{++} \left( \mathcal{G}_k \right) := \left\{ X \in L^{2} \left( \mathcal{G}_k \right)
\big| X > 0 \right\}$, $k=0,...,T.$ Adaptedness of stochastic processes is always meant with respect to $\left( \mathcal{G}_k \right)_{k=0,...,T}$,
unless otherwise stated. We consider arbitrary incomplete no-arbitrage financial markets consisting of $N$
risky securities and one risk-less bond. The price process of each risky asset $i=1,...,N$ is a positive adapted
process $ \left( S^{i}_k \right)_{k=0,...,T}$. Each security $i=1,...,N$ pays a dividend in the next period.
The corresponding divided process is non-negative, adapted and labeled by $ \left( d^{i}_k \right)_{k=1,...,T}$.
The interest rate  process $ \left( r_k \right)_{k=1,...,T} $ representing the risk-less bond is predictable and non-negative.
The payoff space (at each period $k=1,...,T$) is defined by
\[
\mathcal{L}_k := \left\{
\pi^{k-1}_{0} \left( 1 + r_k \right) +
\sum_{i=1}^{N} \pi^{k-1}_{i} \left( S^{i}_k + d^{i}_k \right)
 \big| \pi^{k-1}_i \in L^{2} \left( \mathcal{G}_{k-1} \right), i=0,...,N
\right\} ,
\]
and $\mathcal{L}_0 := \{ 0 \}.$ Note that
$ L^{2} \left( \mathcal{G}_{k-1} \right) \subseteq \mathcal{L}_k \subseteq L^{2} \left(
\mathcal{G}_{k}  \right) $, for all $k=1,...,T.$ We denote by $P^{\mathcal{L}}_k : L^{2} \left( \mathcal{G}_T \right) \to \mathcal{L}_k$, $k=1,...,T,$
the orthogonal projection of the space
$L^{2} \left( \mathcal{G}_T \right)$ onto the subspace $\mathcal{L}_k$. As shown in Lemma 2.5 in Malamud and Trubowitz \cite{MT},
there exists a unique (normalized) state price density (SPD) $\left( M_k \right)_{k=0,...,T}$, which is associated with the wealth
spaces $\left( \mathcal{L}_k \right)_{k=1,...,T}$: $M_0=1$,
\begin{equation}
 S^{i}_{k-1} M_{k-1}  = E \left[  \left( S^{i}_{k} + d^{i}_k \right) M_k \big| \mathcal{G}_{k-1} \right],
\label{aggregate_SPD}
\end{equation}
for all $i=1,...,N$,
\begin{equation}
 M_{k-1}  = E \left[  \left( 1 + r_k \right) M_k \big| \mathcal{G}_{k-1} \right] ,
\label{aggregate_SPD2}
\end{equation}
for all $k=1,...,T$, and $M_k \in \mathcal{L}_k,$ $k=1,...,T.$ This process is referred to as the \textit{aggregate SPD}. Generally speaking,
the aggregate SPD can take non-positive values (see the discussion after Lemma 2.5 in Malamud and Trubowitz \cite{MT}). For simplicity,
we consider only markets with a non-vanishing aggregate SPD. The decision maker in our model is trading in the market and aiming to maximize
his habit-forming preference functional. The endowment stream $\left( \e_k \right)_{k=0,...,T}$ of the agent is non-negative and adapted.
A feasible consumption stream is a non-negative adapted process $\left( c_k \right)_{k=0,...,T}$ of the form
\begin{equation}
c_k = \e_k + W_k - E \left[ \frac{M_{k+1} }{ M_k } W_{k+1}
\big| \mathcal{G}_k \right] ,
\label{consumption_def}
\end{equation}
where $W_k \in \mathcal{L}_k,$ $k=0,...,T$, and $W_{T+1}=0.$ Here,
the processes $\left( W_k \right)_{k=1,...,T}$ and $ \left( E \left[ \frac{M_{k+1} }{ M_k } W_{k+1}
\big| \mathcal{G}_k \right] \right)_{k=0,...,T-1}$ can be interpreted as the wealth and
investment of the investor respectively. The corresponding utility maximization problem is:
\begin{equation}
\sup_{ \left( c_k \right)_{k=0,...,T} \in \mathcal{B} } \sum_{k=0}^{T}
e^{- \rho k }
E \left[ \frac{ \left( c_{k} - \sum_{l=0}^{k-1} \beta^{(k)}_l c_l \right)^{1-\gamma } }{1 - \gamma } \right] ,
\label{utility_maximizaiton}
\end{equation}
where $\mathcal{B}$ is the set of all feasible consumption policies  $\left( c_k \right)_{k=0,...,T}$ satisfying
the constraint $c_{k} \geq \sum_{l=0}^{k-1} \beta^{(k)}_l c_l,$
for all $k=1,...,T.$ The non-negative constants $ \beta^{(k)}_l $, $k=0,...,T,$ $l=0,...,k-1$,
measure the impact of the habit-formation affect on the individual. The constants $\rho$ and
$\gamma$ are viewed as the impatience and risk-aversion coefficients respectively.
Theorem 2.3 in Muraviev \cite{MUR} guarantees that there exists a unique strictly positive optimal consumption stream
$\left( c^{*}_k \right)_{k=0,...,T}$ solving to the optimization problem (\ref{utility_maximizaiton}). We denote by
\[
\widetilde{M}_{k} =  M_k +  \sum_{ l = k+1 }^{T} \
\sum_{j=1}^{l-k} \ \sum_{ k \leq s_{j} < ... < s_1 < l } \beta^{ (
l )  }_{s_1} \beta^{(s_1)}_{s_2} ... \beta^{(s_{j})}_{k} E \left[
M_l  \big| G_k \right]  ,
\]
for $k=0,...,T$, the \textit{perturbed aggregate SPD}. We introduce now the following classes of financial markets.

\begin{Definition}[Malamud and Trubowitz \cite{MT}] An incomplete market is said to be of class $\mathcal{C}$, if
there exists an intermediate filtration $\left( \mathcal{H}_k
\right)_{ k=1,...,T}$ such that
\[
 G_{k-1} \subseteq \mathcal{H}_k
\subseteq \mathcal{G}_k,
\]
and $ P^{k}_{\mathcal{L}} \big[ \cdot \big] = E
\big[ \cdot  | \mathcal{H}_t \big],$ for all $ k=1, ... , T $.
\label{marketc}
\end{Definition}

\begin{Definition}
A financial market is called idiosyncratically incomplete, if there
exist two filtrations $( \mathcal{F}_k )_{ k = 0 , ... ,  T}$ and
$( \mathcal{G}_k )_{ k = 0 , ... ,  T}$ such that:
\newline
(i) $ \mathcal{F}_0 = \mathcal{G}_0 = \{ \emptyset , \Omega \}$, and
$ \mathcal{F}_k \subseteq \mathcal{G}_k $, for all $ k = 1 , ... , T.$
\newline
(ii) The market is complete with respect to $ ( \mathcal{F}_k )_{ k = 0 , ... ,  T} $, and the endowment
stream $( \epsilon_k )_{ k = 0 , ... , T } $ is adapted to $( \mathcal{G}_k )_{ k = 0 , ... ,  T}$.
\newline
(iii) For each $k=0,...,T-1,$ and an arbitrary random variable $ X \in L^{2} \left( \mathcal{F}_{ k + 1 } \right) $, we have
\[
E \left[ X | \mathcal{G}_k \right] = E \left[ X | \mathcal{F}_k \right].
\]
\label{def1}
\end{Definition}

We now state the following results.

\begin{Theorem}
We have
\begin{equation}
 P^{ k }_{ \mathcal{L} } \left[ \frac{ R^{*}_{k}  }{ R^{*}_{k - 1} } \right] =  \frac{ M_{k} }{ M_{k-1}
 },
\label{new_eq0}
\end{equation}
for $ k = 1 , ... , T $, where
\[
R^{*}_{k}  :=   e^{ - \rho k } \left( c^{*}_k - \sum_{j=0}^{k-1} \beta^{(k)}_j c^{*}_j \right)^{-\gamma}
- \sum_{ m = k + 1}^{T} \beta^{(m)}_{ k }  e^{ - \rho m } E \left[
\left(   c^{*}_m  - \sum_{j=0}^{m-1} \beta^{(m)}_j c^{*}_j \right)^{-\gamma}  \bigg| \mathcal{G}_{ k }
\right] ,
\]
for $k=0,...,T$, is a positive SPD.
\label{new0}
\end{Theorem}
${}$
\newline
\textbf{Proof of Theorem \ref{new0}.} See the proof of Theorem 2.3 in Muraviev \cite{MUR}. $\qed$
\newline
\newline
The preceding statement admits a simplified form in the setting of some concrete markets.

\begin{Theorem}
For arbitrary incomplete markets with a deterministic interest rate, or for idiosyncratically incomplete markets, we have
\begin{equation}
P^{\mathcal{L}}_{k} \left[
\left( c^{*}_{k} - \sum_{ l = 0 }^{ k-1} \beta^{(k)}_l c^{*}_l \right)^{-\gamma}
 \right]
=
e^{\rho} \frac{\widetilde{M}_{k}}{\widetilde{M}_{k-1}}
\left( c^{*}_{k-1} - \sum_{ l = 0 }^{ k-2} \beta^{(k-1)}_l c^{*}_l \right)^{-\gamma},
\label{first_order}
\end{equation}
for all $k=1,...,T.$
\label{opt_consumption}
\end{Theorem}

\textbf{Proof of Theorem \ref{opt_consumption}.} See the proof of Theorem 4.1 in Muraviev \cite{MUR}. $\qed$



\section{Estimates}

In the present section we provide estimates for the optimal consumption stream and wealth process in terms of the individual's
\textit{endowments}, \textit{risk-aversion}, \textit{impatience rates}, \textit{degree of habits} and the \textit{aggregate SPD}.
We first set some notation and then introduce an auxiliary lemma.

\begin{Definition}
Let $(X_k)_{k=1,...,T}$ be an adapted process. The upper hedging price of the process
$(X_k)_{k=1,...,T}$ is defined as the minimal number $X^{u}_0 \in R$ such that there exists a wealth process
$W_k \in \mathcal{L}_k,$ $k=1,...,T,$ $W_{T+1}=0,$ which satisfies:
$$
W_k - E \left[  \frac{M_{k+1}}{M_k} W_{k+1} \big| \mathcal{G}_k \right]  \geq X_k,
$$
for all $k=1,...,T$, and
$$
E \left[ M_1 W_1 \right] \leq X^{u}_0.
$$
\end{Definition}

\begin{Lemma}
Consider a market of type $\mathcal{C}.$ For an adapted process $(X_k)_{k=1,...,T}$, set
$X^{u}_T := esssup \left[ X_T \big| \mathcal{H}_{T} \right]$,
\begin{equation}
X^{u}_k := esssup \left[ X_k + E \left[ \frac{M_k}{M_{k-1}} X^{u}_{k+1} \big| \mathcal{G}_{k} \right]  \big| \mathcal{H}_k \right],
\label{suph1}
\end{equation}
for all $k=1,...,T-1,$ and
\begin{equation}
X^{u}_0 := E \left[ M_1 X^{u}_1 \right].
\label{suph2}
\end{equation}
Then, $X^{u}_0$ is the upper hedging price of the process $(X_k)_{k=1,...,T}$.
The process $(X^{u}_k)_{k=0,...,T}$ is referred to as the upper hedging wealth process.
\label{MTW}
\end{Lemma}

\textbf{Proof of Lemma \ref{MTW}.} See the proof of Proposition 2.17 in
Malamud, Trubowitz and W\"uthrich \cite{MTW}. $\qed$
\newline
\newline
We provide below estimates for the optimal consumption stream.

\begin{Theorem}
Let $\left( \widehat{c}_k \right)_{k=0,...,T}$ and $( \widehat{W}_k )_{k=1,...,T}$ denote the optimal consumption stream and
wealth process respectively of an investor trading either in an idiosyncratically incomplete market, or in a market of
type $\mathcal{C}$ with a deterministic interest rate (i.e., $(r_k)_{k=1,...,T}$ are non-negative constants), and solving the utility
maximization problem (\ref{utility_maximizaiton}). Then, we have
\begin{equation}
\left( (-\e)^{U}_k  + \widehat{W}_k \right) m_k + \sum_{j=0}^{k-1} \xi^{k}_j \widehat{c}_j  \leq \widehat{c}_k \leq \left( \e^{U}_k  + \widehat{W}_k \right) m_k + \sum_{j=0}^{k-1} \xi^{k}_j \widehat{c}_j ,
\label{2}
\end{equation}
for all $k=1,...,T,$
\begin{equation}
\left( (-\e)^{U}_0 + \e_0 \right) m_0 \leq \widehat{c}_0 \leq \left( \e^{U}_0 + \e_0 \right) m_0,
\label{3}
\end{equation}
and
\begin{equation}
\sum_{j=0}^{k-1} \alpha^{k}_{j} \widehat{c}_j - \e^U_k
\leq
\widehat{W}_{k} \leq \sum_{j=0}^{k-1} \alpha^{k}_{j} \widehat{c}_j - (-\e)^U_k,
\label{4}
\end{equation}
for all $k=1,...,T.$ Here, $(\xi^{n}_l)_{l=1,...,T;n=0,...,l-1}$,$(\alpha^{n}_l)_{l=1,...,T;n=0,...,l-1}$ and $(m_l)_{l=0,...,T}$ are given explicitly
in (\ref{1'}), (\ref{3'}), (\ref{4'}), (\ref{6'}), (\ref{7'}), (\ref{8'}) and (\ref{9'}); the upper hedging wealth processes $(\e^{U}_k)_{k=0,...,T}$
and $((-\e)^{U}_k)_{k=0,...,T}$ corresponding to $(\e_k)_{k=1,...,T}$ and $(-\e_k)_{k=1,...,T}$ respectively, are given in Lemma \ref{MTW}.
\label{estimates_habits}
\end{Theorem}

\textbf{Proof of Theorem \ref{estimates_habits}.} The proof is by backward induction. First, observe that by using (\ref{consumption_def}) for $k=T$ and the fact that $\widehat{c}_T \geq 0,$ we get that (\ref{2}) for $k=T$ is satisfied with
\begin{equation}
m_T := 1 \ \ \ ; \ \ \ \xi^{T}_j = 0,
\label{1'}
\end{equation}
for all $j=0,...,T-1$, and
\begin{equation}
\eta_T :=  essinf \left[ \e_T \big| \mathcal{H}_T \right] \ \ \ ; \ \ \ \eta'_T := esssup \left[ \e_T \big| \mathcal{H}_T \right].
\label{2'}
\end{equation}
An application of (\ref{2}) for $k=T$ on (\ref{first_order}) yields
\[
E \left[ \left( \widehat{W}_T + esssup \left[ \e_T \big| \mathcal{H}_T \right]
- \sum_{j=0}^{T-1} \beta^{(T)}_j \widehat{c}_j \right)^{-\gamma} \big| \mathcal{H}_T  \right]
\]
\[
\leq \frac{\widetilde{M}_{T}}{\widetilde{M}_{T-1}} e^{\rho} \left( \widehat{c}_{T-1} - \sum_{j=0}^{T-2} \beta^{(T-1)}_j \widehat{c}_j  \right)^{ - \gamma }
\]
\[
\leq E \left[ \left( \widehat{W}_T + essinf \left[ \e_T \big| \mathcal{H}_T \right]
- \sum_{j=0}^{T-1} \beta^{(T)}_j \widehat{c}_j \right)^{-\gamma} \big| \mathcal{H}_T  \right].
\]
Since the expressions above within the the conditional expectations are $\mathcal{H_T}-$measurable, we obtain
\begin{equation}
\sum_{j=0}^{T-1} \alpha^{T}_{j} \widehat{c}_j + \delta_{T}
\leq
\widehat{W}_{T} \leq \sum_{j=0}^{T-1} \alpha^{T}_{j} \widehat{c}_j + \delta'_{T},
\label{a}
\end{equation}
with
\begin{equation}
\alpha^T_{T-1} :=  e^{ - \frac{ \rho }{ \gamma } } \left( \frac{\widetilde{M}_T}{\widetilde{M}_{T-1}} \right)^{-1 / \gamma}  + \beta^{(T)}_{T-1},
\label{3'}
\end{equation}
\begin{equation}
\alpha^T_{j} := \beta^{(T)}_j - \beta_{j}^{(T-1)} e^{ - \frac{ \rho }{ \gamma } } \left( \frac{\widetilde{M}_T}{\widetilde{M}_{T-1}} \right)^{-1 / \gamma},
\label{4'}
\end{equation}
for $j=0,...,T-2,$ and
\begin{equation}
\delta_T := - esssup \left[ \e_T \big| \mathcal{H}_T \right]
\ \ \ ; \ \ \
\delta'_T := - essinf \left[ \e_T \big| \mathcal{H}_T \right].
\label{5'}
\end{equation}
Assume now that
\[
\sum_{j=0}^{k} \alpha^{k+1}_{j} \widehat{c}_j + \delta_{k+1}
\leq
\widehat{W}_{k+1} \leq \sum_{j=0}^{k} \alpha^{k+1}_{j} \widehat{c}_j + \delta'_{k+1}.
\]
Recall that $\widehat{c}_k = \e_k + \widehat{W}_k - E \left[ \frac{M_{k+1}}{M_k} \widehat{W}_{k+1} \big| \mathcal{G}_k \right] $, and thus we get
\begin{equation}
\eta_k + m_k \widehat{W}_k + \sum_{j=0}^{k-1} \xi^{k}_j \widehat{c}_j  \leq \widehat{c}_k \leq \eta'_k + m_k \widehat{W}_k + \sum_{j=0}^{k-1} \xi^{k}_j \widehat{c}_j,
\label{b'}
\end{equation}
where
\begin{equation}
m_k := \frac{1}{ 1 + E \left[ \alpha^{k+1}_k \frac{M_{k+1}}{M_k} \big| \mathcal{G}_{k } \right] },
\label{6'}
\end{equation}
\begin{equation}
\xi^{k}_j := \frac{ E \left[ \alpha^{k+1}_j \frac{M_{k+1}}{M_k}  \big| \mathcal{G}_{k} \right] }{
1 + E \left[ \alpha^{k+1}_k \frac{M_{k+1}}{M_k}  \big| \mathcal{G}_{k} \right] },
\label{7'}
\end{equation}
for $j=0,...,k-1$, and
\[
\eta'_k := \frac{ esssup \left[ \e_k  -E \left[  \delta_{k+1} \frac{ M_{k+1} }{ M_k } \big| \mathcal{G}_k \right] \big| \mathcal{H}_k \right] }{ 1 + E \left[ \alpha^{k+1}_k \frac{ M_{k+1} }{ M_k } \big| \mathcal{G}_k \right] } \ \ \ ; \ \ \ \eta_k :=
\frac{ essinf \left[ \e_k  -E \left[  \delta'_{k+1} \frac{ M_{k+1} }{ M_k } \big| \mathcal{G}_k \right] \big| \mathcal{H}_k \right] }{ 1 + E \left[ \alpha^{k+1}_k \frac{ M_{k+1} }{ M_k } \big| \mathcal{G}_k \right] },
\]
for all $k=1,...,T$. Next, by combining (\ref{first_order}) with the previous inequality, we obtain
\[
\left( \eta_k + m_k \widehat{W}_k + \sum_{j=0}^{k-1}  \left( \xi^{k}_j - \beta^{(k)}_j \right) \widehat{c}_j \right)^{-\gamma}
\geq
\]
\[
\frac{ \widetilde{M}_k}{\widetilde{M}_{k-1} } e^{\rho} \left( \widehat{c}_{k-1} - \sum_{j=0}^{k-2}  \beta^{(k-1)}_{j} \widehat{c}_j \right)^{- \gamma }
\geq
\]
\[
\left( \eta'_k + m_k \widehat{W}_k + \sum_{j=0}^{k-1}  \left( \xi^{k}_j - \beta^{(k)}_j \right) \widehat{c}_j \right)^{-\gamma} ,
\]
this implies that
\begin{equation}
\sum_{j=0}^{k-1} \alpha^{k}_j \widehat{c}_j + \delta_k
\leq \widehat{W}_k \leq
\sum_{j=0}^{k-1} \alpha^{k}_j \widehat{c}_j + \delta'_k,
\label{c'}
\end{equation}
where
\begin{equation}
\alpha^{k}_{k-1} := \frac{ \left( \frac{ \widetilde{M}_{k} }{ \widetilde{M}_{k-1} } \right)^{ - 1 / \gamma } e^{ - \frac{ \rho }{ \gamma } }
+ \beta^{(k)}_{k-1} - \xi^{k}_{k-1} }{m_k} ,
\label{8'}
\end{equation}
\begin{equation}
\alpha^{k}_{j} := \frac{ \beta^{(k)}_j - \beta^{(k-1)}_j   e^{ - \frac{ \rho }{ \gamma } } \left( \frac{ \widetilde{M}_{k} }{ \widetilde{M}_{k-1} } \right)^{ - 1 / \gamma }
- \xi^{k}_{j} }{m_k} ,
\label{9'}
\end{equation}
\[
\delta_k := - \frac{ \eta'_k }{ m_k } \ \ \ ; \ \ \ \delta'_k := - \frac{ \eta_k }{ m_k },
\]
for all $k=1,...,T$. Finally, we obtain the inequality
\[
m_0 \e_0 + \eta_0 \leq \widehat{c}_0 \leq m_0 \e_0 + \eta'_0 ,
\]
with
$ m_0 := \frac{1}{1 + E \left[ \alpha^{1}_0 M_1 \right] }$, and
\[
\eta_0 := - \frac{ E \left[ M_1 \delta'_1 \right] }{1 + E \left[ \alpha^{1}_0 M_1 \right] }
\ \ \ ; \ \ \
\eta'_0 := - \frac{ E \left[ M_1 \delta_1 \right] }{1 + E \left[ \alpha^{1}_0 M_1 \right] }.
\]
One can verify that $ \delta_k= - \e^U_k$ and $\delta'_k = (-\e)^U_k,$ for all $k=1,...,T.$
This completes the proof. $\qed$
\newline
\newline
The following statement is a simplified version of Theorem \ref{estimates_habits}
for the case where habits are not incorporated.

\begin{Corollary}
Denote by $(\overline{c}_k)_{k=0,...,T}$ and $(\overline{W}_k)_{k=1,...,T}$
the optimal consumption and wealth process of an individual trading in a market of type $\mathcal{C}$ and solving the utility
maximization problem (\ref{utility_maximizaiton}) with no habits, i.e.,
$\beta^{(k)}_{l}=0$, for all $k=1,...,T$ and $l=0,...,k-1.$ Then, under the notations of Theorem \ref{estimates_habits},
the following is satisfied
\begin{equation}
\left( (-\e)^{U}_k  + \overline{W}_k \right) m_k \leq \overline{c}_k \leq \left( \e^{U}_k  + \overline{W}_k \right) m_k ,
\label{2a}
\end{equation}
for all $k=1,...,T,$
\begin{equation}
\left( (-\e)^{U}_0 + \e_0 \right) m_0 \leq \overline{c}_0 \leq \left( \e^{U}_0 + \e_0 \right) m_0,
\label{3a}
\end{equation}
and
\begin{equation}
\alpha^{k}_{k-1} \overline{c}_{k-1} - \e^U_k
\leq
\overline{W}_{k} \leq \alpha^{k}_{k-1} \overline{c}_{k-1} - (-\e)^U_k,
\label{4a}
\end{equation}
for all $k=1,...,T.$
\label{estimates_no_habits}
\end{Corollary}

\textbf{Proof of Corollary \ref{estimates_no_habits}.} The assertion follows immediately from Theorem \ref{estimates_habits} and Proposition 2.9
in Malamud and Trubowitz \cite{MT} (which corresponds to Theorem \ref{new0} with no habits) . $\qed$

\section{Asymptotics}
Denote by $ \left( c_k(\e_0,\e_1,...,\e_T) \right)_{ k=0,...,T}$ and $ \left(W_k(\e_0,\e_1,...,\e_T)\right)_{k=0,...,T}$ the optimal consumption
stream and wealth process, respectively, solving the utility maximization problem (\ref{utility_maximizaiton}).
Note that the scaling property of the power utility function yields
\begin{equation}
 \frac{ c_k(\e_0,\e_1,...,\e_T)}{\e_0} = c_k \left( 1, \frac{\e_1}{\e_0} , ... , \frac{\e_T}{\e_0}  \right),
\label{scaling1}
\end{equation}
for all $k=0,...,T,$ and
\begin{equation}
 \frac{ W_k(\e_0,\e_1,...,\e_T)}{\e_0} = W_k \left( 1, \frac{\e_1}{\e_0} , ... , \frac{\e_T}{\e_0}  \right),
\label{scaling0}
\end{equation}
for all $k=1,...,T$. In the current section, we let $\e_1,...,\e_T$ be fixed, and study the asymptotic behavior (as $\e_0 \to \infty$)
of the quanteties (\ref{scaling1}) and (\ref{scaling0}), for various models:
incomplete markets with a positive aggregate SPD, incomplete markets with a deterministic interest rate and
idiosyncratically incomplete markets. For these cases, we show that the limits corresponding to (\ref{scaling1}) and (\ref{scaling0})
exist and equal to $c_k(1,0,...,0)$ and $W_k(1,0,...,0)$, respectively. Therefore, the problem amounts to checking the
continuity of the functions $c_k \left( 1,  \e_1 , ... , \e_T   \right),$ $k=0,...,T$ and
$W_k \left( 1, \e_1 , ... , \e_T  \right),$ $k=1,...,T$ at $(1,0,...,0).$

\subsection{Positive Aggregate SPD}
First, we consider \textit{arbitrary} incomplete
financial markets with a \textit{positive} aggregate SPD $ \left( M_k\right)_{k=0,...,T}$.
In this setting, we let $( c^{*}_k ( \e_0 , \e_1 ... , \e_T ))_{k=0,...,T}$ and $(W^{*}_k ( \e_0 , \e_1 ... , \e_T ) )_{k=1,...,T}$ denote
the corresponding optimal consumption stream and wealth process respectively. We set further $c^{*}_k= c^{*}_k ( 1, 0 ... , 0 ) $, $k=0,...,T$, and
$W^{*}_k= c^{*}_k ( 1, 0 ... , 0 ) $, $k=1,...,T$. By Theorem \ref{new0} and identity (\ref{consumption_def}), we have
\begin{equation}
c^{*}_k = W^{*}_k - E \left[  \frac{ M_{k+1} }{ M_k }
W^{*}_{k+1}  \big| \mathcal{G}_{k} \right] \ \ ; \ \ c^{*}_0 = 1 - E \left[ M_1 W^{*}_1 \right],
\label{consumption_no_end}
\end{equation}
and
\begin{equation}
P^{ k }_{ \mathcal{L} } \left[ \frac{ R^{*}_{k}  }{ R^{*}_{k - 1} } \right] =  \frac{ M_{k} }{ M_{k-1}
 },
\label{wealth_no_end}
\end{equation}
for all $k=1,...,T$, where
\[
R^{*}_{k}  :=   e^{ - \rho k } \left( c^{*}_k - \sum_{j=0}^{k-1} \beta^{(k)}_j c^{*}_j \right)^{-\gamma}
- \sum_{ m = k + 1}^{T} e^{ - \rho m } \beta^{(m)}_{ k }  E \left[
\left(   c^{*}_m  - \sum_{j=0}^{m-1} \beta^{(m)}_j c^{*}_j \right)^{-\gamma}  \bigg| \mathcal{G}_{ k }
\right] ,
\]
for $k=0,...,T$.
\newline
\newline
We exhibit now the main result of the subsection.

\begin{Theorem}
We have
\begin{equation}
\lim_{ \e_0 \to \infty } \frac{ c^{*}_k ( \e_0 , \e_1 ... , \e_T ) }{ \e_0 } = c^{*}_k,
\label{consumption_asymptotics}
\end{equation}
$P-$a.s., for all $k=0,...,T$, and
\begin{equation}
\lim_{ \e_0 \to \infty } \frac{ W^{*}_k ( \e_0 , \e_1 ... , \e_T ) ( \epsilon_0) }{ \e_0 } = W^{*}_k,
\label{wealth_asymptotics}
\end{equation}
$P-$a.s., for all $k=1,...,T$.
\label{main1}
\end{Theorem}

We first prove the following weaker result.

\begin{Lemma}
We have
\begin{equation}
\limsup_{ \epsilon_0 \to \infty } \frac{c^{*}_k ( \e_0 , \e_1 ... , \e_T ) }{ \e_0 } < \infty \ \ ; \ \
\liminf_{ \epsilon_0 \to \infty } \frac{ c^{*}_k ( \e_0 , \e_1 ... , \e_T ) }{ \e_0 } >
0 ,
\end{equation}
$P-$a.s, for all $k=0,...,T-1$, and
\begin{equation}
\limsup_{ \epsilon_0 \to \infty } \frac{ W^{*}_k ( \e_0 , \e_1 ... , \e_T ) }{ \e_0 } < \infty \ \ ; \ \
\liminf_{ \epsilon_0 \to \infty } \frac{ W^{*}_k ( \e_0 , \e_1 ... , \e_T ) }{ \e_0 } > 0 ,
\end{equation}
$P-$a.s, for all $k=1,...,T$.
\label{limsup}
\end{Lemma}

\textbf{Proof of Lemma \ref{limsup}.}
To simplify notations ($\e_1,...,\e_T$ are fixed), we denote $c_k(\e_0):=c^{*}_k(\e_0, \e_1 , ..., \e_T)$, $k=0,...,T$, and
$W_{k}(\e_0):=W^{*}_{k}(\e_0, \e_1 , ..., \e_T)$, $k=1,...,T$.
First, let us show that $ \liminf_{\e_0 \to \infty } \frac{ W_1 ( \e_0 ) }{ \e_0 } \geq 0, $ $P-$a.s.
Assume that there exists some constant $a_1 < 0 $ such that $P \left(  \liminf_{\e_0 \to \infty } \frac{W_1 ( \e_0 ) }{ \e_0 } < a_1 \right) >0. $
Then, since the optimal consumption stream and $(M_k)_{k=0,...,T}$ are positive, and $ c_1 (\e_0) = \e_1 + W_1( \e_0 ) -
E \left[ \frac{M_2}{M_1} W_2 ( \e_0) \big| \mathcal{G}_1 \right]  > 0,$ it follows that $P \left(  \liminf_{\e_0 \to \infty } \frac{W_2 ( \e_0 ) }{ \e_0 } < a_2 \right) > 0, $
for some $a_2 < 0.$ Continuing inductively, we obtain that $P \left(  \liminf_{\e_0 \to \infty } \frac{W_T ( \e_0 ) }{ \e_0 } < a_T \right) > 0, $
for some constant $a_T < 0$. This is a contradiction since $c_T (\e_0)$ is positive and $c_t( \e_0) = \e_T + W_T(\e_0).$
In the same way, one checks that $ \liminf_{\e_0 \to \infty } \frac{W_k ( \e_0 ) }{ \e_0 } \geq 0, $ $P-$a.s, $k=1,...,T.$
We treat the upper limits now.
We have
$\limsup_{ \e_0 \to \infty }  E \big[ M_1 \times $ $\frac{W_1 ( \e_0) }{\e_0} \big] =
\limsup_{\e_0 \to \infty} \left( 1 - \frac{c_0 ( \e_0) }{ \e_0 } \right) \leq 1.$
Since $M_1$ is a positive random variable, it follows that $\limsup_{\e_0 \to \infty} \frac{W_{1} (\e_0) }{ \e_0 } < \infty,$
$P-$a.s. The identity $ \frac{c_k(\e_0)}{\e_0} =
\frac{\e_k}{\e_0} + \frac{W_k}{ \e_0 } - E \left[ \frac{M_{k+1}}{M_k}  \frac{W_{k+1} (\e_0 ) }{\e_0} \big| \mathcal{G}_k \right] ,$
for $k=1,...,T$, the fact that $ \liminf_{\e_0 \to \infty } \frac{W_k ( \e_0 ) }{ \e_0 } \geq 0, $ $P-$a.s, $k=1,...,T,$
and the positivity of the process $(M_k)_{k=0,...,T}$ conclude the proof for the upper limits.
Next, we treat the lower limits. First, we claim that $\liminf_{\e_0 \to \infty } \frac{c_1 ( \e_0 )}{ c_0 ( \e_0) } > \beta^{(1)}_0.$
Assume on the contrary that this is not the case, and multiply the first order condition (\ref{new_eq0}) for $k=1$ by $c^{\gamma}_0(\e_0)$:
\[
P^{1}_{\mathcal{L}} \left[ \left( \frac{c_1(\e_0) }{ c_0(\e_0) } - \beta^{1}_0 \right)^{-\gamma}
- \sum_{ j = 2}^{T} \beta^{(j)}_{ 1 }
E \left[
\left( \frac{c_j(\e_0)}{c_0(\e_0)} - \sum_{l=0}^{j-1} \beta^{(j)}_l \frac{c_l (\e_0)}{c_0(\e_0)} \right)^{-\gamma}
\bigg| \mathcal{G}_1
\right]
\right] =
\]
\[
e^{\rho} M_1
\left( 1
- \sum_{ j = 1}^{T} \beta^{(j)}_{ 0 } E \left[ \left( \frac{c_j(\e_0)}{c_0(\e_0)} - \sum_{l=0}^{j-1} \beta^{(j)}_l
 \frac{c_l (\e_0)}{c_0(\e_0)}  \right)^{-\gamma} \right]
\right).
\]
By Theorem \ref{new0} (recall that $(R^{*})_{k=0,...,T}$ is a positive SPD), we have
\[
0 \leq \sum_{ j = 1}^{T} \beta^{(j)}_{ 0 } E \left[ \left( \frac{c_j(\e_0)}{c_0(\e_0)} - \sum_{l=0}^{j-1} \beta^{(j)}_l
 \frac{c_l (\e_0)}{c_0(\e_0)}  \right)^{-\gamma} \right] < 1 .
\]
Therefore, we get a contradiction by applying an expectation on both sides of the equation, and observing that
\[
\limsup_{\e_0 \to \infty}
E \left[ \left( \frac{c_1(\e_0) }{ c_0(\e_0) } - \beta^{1}_0 \right)^{-\gamma}
- \sum_{ j = 2}^{T} \beta^{(j)}_{ 1 }
E \left[
\left( \frac{c_j(\e_0)}{c_0(\e_0)} - \sum_{l=0}^{j-1} \beta^{(j)}_l \frac{c_l (\e_0)}{c_0(\e_0)} \right)^{-\gamma}
\bigg| \mathcal{G}_1
\right]
\right] = \infty,
\]
whereas the right hand side is bounded. In the same manner, one checks that
\begin{equation}
\liminf_{\e_0 \to \infty }  \frac{ c_{k} ( \e_0 )}{ c_0 ( \e_0 )}  > \alpha_k,
\label{liminf}
\end{equation}
for all $k=1,...,T$, where
\begin{equation}
\alpha_k := \sum_{l=0}^{k-1} \prod_{l \geq i_k > ... > i_1 > 0} \beta^{(k)}_l \beta^{(l)}_{i_k} ...  \beta^{(i_1)}_{0} \geq 0.
\label{alpha}
\end{equation}
Recall now that $c_T ( \e_0 ) = \e_T + W_T ( \e_0 ) $, hence
$$
\liminf_{\e_0 \to \infty } \frac{ W_T ( \e_0 ) }{ c_0 ( \e_0 ) } = \liminf_{\e_0 \to \infty } \frac{ c_T ( \e_0 ) }{ c_0 ( \e_0 ) }  >
\alpha_T,
$$
$P-$a.s. By (\ref{consumption_def}) for $k=T-1 $, we get
\[
 \frac{ c_{T-1} ( \e_0 ) }{ c_0 ( \e_0 ) } = \frac{ \e_{T-1}  }{ c_0 ( \e_0 ) } + \frac{ W_{T-1} ( \e_0 ) }{ c_0 ( \e_0 ) }
- E \left[ \frac{M_T}{ M_{T-1} } \frac{ W_{T} ( \e_0 ) }{ c_0 ( \e_0 ) }  | \mathcal{G}_{T-1} \right],
\]
$P-$a.s., and we conclude that
$$
\liminf_{\e_0 \to \infty } \frac{ W_{T-1} ( \e_0 ) }{ c_0 ( \e_0 ) }  >
\alpha_T  E \left[ \frac{M_T }{ M_{T-1} } \big| \mathcal{G}_{T-1} \right] - \limsup_{\e_0 \to \infty}
\frac{ \e_{T-1}  }{ c_0 ( \e_0 ) },
$$
$P-$a.s. In the same manner, one can verify that
\begin{equation}
\liminf_{\e_0 \to \infty } \frac{ W_{k} ( \e_0 ) }{ c_0 ( \e_0 ) }  >
- \limsup_{ \e_0 \to \infty} \frac{ \sum_{j=k}^{T-1} \e_j }{ c_0 ( \e_0 ) }
+ \alpha_T E \left[ \frac{M_T}{M_k} \big| \mathcal{G}_k \right]  ,
\label{eq_1}
\end{equation}
$P-$a.s., for all $k=1,...,T-1.$ Now, assume on the contrary that $\liminf_{ \e_0 \to \infty} \frac{ c_0 (\e_0 ) }{ \e_0 } = 0.$
Observe that we can rewrite (\ref{consumption_def}) for $k=0$ as
\[
1 =  \frac{ \e_0 }{ c_0 ( \e_0 ) } - E \left[ M_1 \frac{W_1 \left( \e_0 \right) }{c_0 ( \e_0 ) } \right] ,
\]
hence (since $\limsup_{ \e_0 \to \infty } \frac{W_1 \left( \e_0 \right) }{c_0 ( \e_0 ) } < \infty  $), we get a contradiction.
In particular, $\lim_{\e_0 \to \infty } c_0 ( \e_0)
= \infty,$ and thus inequality (\ref{eq_1}) becomes $ \liminf_{\e_0 \to \infty } \frac{ W_{k} ( \e_0 ) }{ c_0 ( \e_0 ) }  >
\alpha_T E \left[ \frac{M_T}{M_k} \big| \mathcal{G}_k \right],$ $P-$a.s., for all $k=1,...,T.$
The proof is now accomplished by noting that the preceding
observations applied on (\ref{liminf}) and (\ref{eq_1}) yield
$$
\liminf_{ \e_0 \to \infty} \frac{c_k ( \e_0) }{ \e_0 } >
\alpha_k \liminf_{ \e_0 \to \infty} \frac{ c_0 ( \e_0) }{ \e_0  } \geq 0 ,
$$
and
$$ \liminf_{ \e_0 \to \infty} \frac{ W_k ( \e_0) }{ \e_0 } >
\alpha_T E \left[ \frac{M_T}{M_k} \big| \mathcal{G}_k \right] \liminf_{ \e_0 \to \infty} \frac{ c_0 ( \e_0) }{ \e_0  } \geq 0
,$$
$P-$a.s., for all $k=1,...,T.$ $\qed$

\begin{Corollary}
We have
\[
\lim_{ \e_0 \to \infty} c_k ( \e_0 )  = \infty,
\]
$P-$a.s., for all $k=0,...,T,$ and
\[
\lim_{\e_0 \to \infty} W_k ( \e_0) = \infty ,
\]
$P-$a.s., for all $k=1,...,T$. $\qed$
\label{cor1}
\end{Corollary}

\textbf{Proof of Corollary \ref{cor1}.} The assertion follows from the lower limits established in
Lemma \ref{limsup}. $\qed$
\newline
\newline
We are now ready to prove the main result of the subsection.
\newline
\newline
\textbf{Proof of Theorem \ref{main1}.} By Lemma \ref{limsup}, there exist two sequences
$ ( \e^{n}_1 )_{n \in N }$ and $ ( {\e}^{n}_2 )_{n \in N }$ of real numbers tending to $+\infty$ such
that $ \lim_{n \to \infty } \frac{ c_k ( {\e}^{n}_i ) }{ {\e}^{n}_i } = c^{(i)}_k $, $P-$a.s., for $k=0,...,T,$
and $ \lim_{n \to \infty } \frac{ W_k ( {\e}^{n}_i ) }{ {\e}^{n}_i } = W^{(i)}_k $, $P-$a.s., for $k=1,...,T$, and $i=1,2,$
where $ 0 < c^{(i)}_k < \infty,$ $P-$a.s., for all $k=0,...,T$, and $ 0 < W^{(i)}_k < \infty,$ $P-$a.s., for all $k=1,...,T$,
and $i=1,2.$ Now, by multiplying equations (\ref{new_eq0}) and (\ref{consumption_def}) by $ ( {\e}^{n}_i)^{\gamma} $ and $ ( {\e}^{n}_i)^{-1} $
respectively, and then letting $n \to \infty$,
we obtain the following identities:
\[
P_{ \mathcal{L} }^{k} \left[ R^{i}_k  \right]
= \frac{ M_{k+1} }{ M_{k} } 
R^{i}_{k-1},
\]
for all $k=0,...,T-1,$ $i=1,2,$ where
\[
R^{i}_{k}  :=   e^{ - \rho k } \left( c^{i}_k - \sum_{j=0}^{k-1} \beta^{(k)}_j c^{i}_j \right)^{-\gamma}
- \sum_{ m = k + 1}^{T} \beta^{(m)}_{ k }  E \left[
\left(   c^{i}_m  - \sum_{j=0}^{m-1} e^{ - \rho m } \beta^{(m)}_j c^{i}_j \right)^{-\gamma}  \bigg| \mathcal{G}_{ k }
\right] ,
 \]
and
\[
c^{(i)}_k = W^{(i)}_k - E \left[ \frac{ M_{k+1} }{ M_k } W^{(i)}_{k+1} \big| \mathcal{G}_{k}  \right]
\]
for all $k=1,...,T$, $i=1,2,$ and
\[
c^{(i)}_0 = 1 - E \left[ M_1 W^{(i)}_1 \right],
\]
for $i=1,2$. Note that the above system of equations corresponds to the solution of the utility maximization $(\ref{utility_maximizaiton})$ with the
endowments $\e_0=1$ and $\e_k=0$ for all $k=1,...,T$ (see (\ref{consumption_no_end}) and (\ref{wealth_no_end})). Therefore,
the uniqueness of the optimal consumption stream and the wealth process implies that: $\lim_{ \e_0 \to \infty } \frac{c_k ( \e_0 ) }{ \e_0 }$
exists, $P-$a.s., for each $k=0,...,T;$ $\lim_{ \e_0 \to \infty } \frac{W_k ( \e_0 ) }{ \e_0 }$ exists, $P-$a.s., for each $k=1,...,T;$
$c^{(1)}_k = c^{(2)}_k = c^{*}_k$ for all $k=0,...,T,$
and $W^{(1)}_k = W^{(2)}_k =W^{*}_k$ for all $k=1,...,T,$ completing the proof of Theorem \ref{main1}. $\qed$

\subsection{Idiosyncratic Incompleteness and Deterministic Interest Rate}

The scaling property of the power utility function combined with Theorem 4.1 in Muraviev \cite{MUR} and Theorem 2.14 in 
Malamud and Trubowitz \cite{MT} simplifies substantially the analysis of the asymptotic behavior of the quantities 
(\ref{scaling1}) and (\ref{scaling0}), for idiosyncratically
incomplete markets and markets of type $\mathcal{C}$ with habits, and arbitrary incomplete markets with no habits. 
Furthermore, it allows us to establish the convergence rates.
Denote by $ \left( \widehat{c}_k( \e_0,\e_1,...,\e_T ) \right)_{k=0,...,T}$ and $ ( \widehat{W}_k( \e_0,\e_1,...,\e_T ))_{k=1,...,T}$
($\left( \underline{c}_k( \e_0,\e_1,...,\e_T ) \right)_{k=0,...,T}$ and $ ( \underline{W}_k( \e_0,\e_1,...,\e_T ))_{k=1,...,T}$) the optimal
consumption and wealth process respectively of an investor solving the utility maximization problem (\ref{utility_maximizaiton}) in an
incomplete market with a deterministic interest rate (idiosyncratically incomplete market).
We set further $\widehat{c}_k := \widehat{c}_k( 1,0,...,0 ) $, $\underline{c}_k := \underline{c}_k( 1,0,...,0 ) $, $k=0,...,T,$
and $\widehat{W}_k := \widehat{W}_k( 1,0,...,0 ) $, $\underline{W}_k := \underline{W}_k( 1,0,...,0 ) $, $k=1,...,T.$
By Theorem \ref{opt_consumption}, we have
\begin{equation}
\widehat{c}_k = \sum_{j=0}^{k-1} \beta^{(k)}_j \widehat{c}_j + \left(
\frac{ \widetilde{M}_k }{ \widetilde{M}_{k-1}} \right)^{- \frac{1}{\gamma}} \e^{- \frac{\rho}{\gamma}} \widehat{c}_0,
\label{classC1}
\end{equation}
for all $k=1,...,T,$
\begin{equation}
\widehat{c}_0 + \sum_{k=1}^{T} E \left[ \frac{M_k}{M_{k-1}} \widehat{c}_k \right] = 1 ,
\label{classC2}
\end{equation}
\begin{equation}
\widehat{W}_k = \sum_{j=k}^{T}  E  \left[ \frac{M_j}{M_k} \widehat{c}_j \big| \mathcal{G}_k \right] ,
\label{classC3}
\end{equation}
for all $k=1,...,T,$ and
\begin{equation}
\underline{c}_k = \underline{W}_k - E \left[ \frac{M_{k+1}}{M_k} \underline{W}_{k+1} \big| \mathcal{G}_k \right]  \ \ ; \ \
\underline{c}_0 = 1 - E \left[ M_1 \underline{W}_1 \right],
\label{class'}
\end{equation}
and
\begin{equation}
P_{\mathcal{L}}^{k} \left[
\left( \underline{c}_k - \sum_{j=0}^{k-1} \beta^{(k)}_j \underline{c}_j  \right)^{-\gamma}  \right] =
\frac{ \widetilde{M}_k }{ \widetilde{M}_{k-1} } e^{\rho}
\left( \underline{c}_{k-1} - \sum_{j=0}^{k-2} \beta^{(k-1)}_j \underline{c}_j \right)^{-\gamma},
\label{class''}
\end{equation}
for all $k=1,...,T$.

\begin{Theorem}
The processes  $(\widehat{c}_k( \e_0,\e_1,...,\e_T ))_{k=0,...,T}$, $(\widehat{W}_k( \e_0,\e_1,...,\e_T ))_{k=1,...,T}$,
$(\underline{c}_k( \e_0,\e_1,...,\e_T ))_{k=0,...,T}$ and $(\underline{W}_k( \e_0,\e_1,...,\e_T ))_{k=1,...,T}$ are $C^{\infty}-$differentiable
with respect to each variable $\e_j,$ $j=1,...,T$. In particular, we have
\begin{equation}
\left| \frac{ \widehat{c}_k ( \epsilon_0) }{ \e_0 } - \widehat{c}_k \right| = O \left( \frac{1}{\e_0} \right),
\label{consumption_asymptotics1'}
\end{equation}
\begin{equation}
\left| \frac{ \underline{c}_k ( \epsilon_0) }{ \e_0 } - \underline{c}_k \right| = O \left( \frac{1}{\e_0} \right),
\label{consumption_asymptotics1}
\end{equation}
as $\e_0 \to \infty$, $P-$a.s., for all $k=0,...,T$, and
\begin{equation}
\left| \frac{ \widehat{W}_k ( \epsilon_0) }{ \e_0 }  - \widehat{W}_k \right| =
O \left( \frac{1}{\e_0} \right),
\label{wealth_asymptotics1}
\end{equation}
\begin{equation}
\left| \frac{ \underline{W}_k ( \epsilon_0) }{ \e_0 }  - \underline{W}_k \right| =
O \left( \frac{1}{\e_0} \right),
\label{wealth_asymptotics1}
\end{equation}
as $\e_0 \to \infty$, $P-$a.s., for all $k=1,...,T$.
\label{main2}
\end{Theorem}

\textbf{Proof of Theorem \ref{main2}.}
The differentiability follows the same ideas as those in the proof of Theorem 4.1 in Muraviev \cite{MUR}, based on the
implicit function theorem, and thus is omitted. The rates of convergence follows directly from
differentiability, (\ref{scaling1}) and (\ref{scaling0}). $\qed$
\newline
\newline
Consider the utility maximization problem (\ref{utility_maximizaiton}) with no habits, i.e.,
$\beta^{(k)}_l=0$, $k=1, ... , T$,  $l=0,...,k-1,$. Let $( c'_k (\e_0, \e_1, ... , \e_T ) )_{k=0,...,T}$
and $(W'_k (\e_0, \e_1, ... , \e_T ) )_{k=1,...,T}$ denote the corresponding optimal consumption and investment processes
respectively. We set further $c'_k := c'_k ( 1, 0 ... , 0 ) ,$ $k=0,...,T$ and $W'_k := W'_k ( 1, 0 ... , 0 ) $, $k=1,...,T.$
By Proposition 2.9 in Malamud and Trubowitz \cite{MT} (which coincides with Theorem \ref{new0} with no habits), we have
\[
P^{k}_{\mathcal{L}} \left[ ( c'_k )^{-\gamma } \right] = e^{\rho} \frac{M_k}{M_{k-1}} \left(c'_{k-1} \right)^{-\gamma},
\]
\[
c'_k = W'_k - E \left[ \frac{M_{k+1}}{M_k} W'_{k+1} \right],
\]
for $k=1,...,T$. If the market is of type $\mathcal{C}$, we get
\[
c'_k = e^{ - \frac{ \rho }{ \gamma } } \left( M_k  \right)^{ - \frac{1}{\gamma}} c'_{0},
\]
\[
W'_k = \sum_{j=k}^{T} E \left[ \frac{M_j}{M_k} c'_k \big| \mathcal{G}_k  \right] .
\]
for $k=1,...,T,$ and
\[
c'_0 + \sum_{j=0}^{T} E \left[ M_k c'_k \right]
= \e'_0 + \sum_{j=0}^{T} E \left[ M_k \e'_k \right].
\]

\begin{Theorem}
The processes $( c'_k (\e_0, \e_1, ... , \e_T ) )_{k=0,...,T}$
and $(W'_k (\e_0, \e_1, ... , \e_T ) )_{k=1,...,T}$ are $C^{\infty}-$differentiable with respect to each $\e_k$, $k=1,...,T.$
In particular, we have
\begin{equation}
\left| \frac{ c'_k ( \epsilon_0 , \e_1 , ... ,\e_T ) }{ \e_0 } - c'_k \right| = O \left( \frac{1}{\e_0} \right),
\end{equation}
for all $k=0,...,T,$ and
\begin{equation}
\left| \frac{ W'_k ( \epsilon_0 , \e_1 , ... ,\e_T ) }{ \e_0 } - W'_k \right| = O \left( \frac{1}{\e_0} \right).
\end{equation}
\label{main3}
\end{Theorem}

\textbf{Proof of Theorem \ref{main3}.} The assertion follows from
Theorem 2.14 in Malamud and Trubowitz \cite{MT}. $\qed$

\section{Equilibrium}

We consider throughout the section preferences with static type of habits that are assigned according to a last period consumption.
Namely, $\beta^{(k)}_{k-1} = \beta \geq 0 ,$ $k=1,...,T$, and $\beta^{(k)}_j=0$, $k=1,...,T,$ $j=0,...,k-2.$ Furthermore, we restrict
our analysis to \textit{complete markets}, i.e., there exists a unique (normalized) positive SPD (which coincides with the aggregate SPD) denoted by
$\left( M_k \right)_{k=0,...,T} $.
We consider an economy inhabited by $N$ (types of) economic investors labeled by $i=1,...,N.$ In accordance with the above framework,
each individual $i$ solves the utility maximization problem:
\begin{equation}
\max_{ c^{i}_0 , ... , c^{i}_T } \sum_{ k = 0 }^{ T } e^{ - \rho_i k }
E \left[ \frac{\left( c^{i}_k - \beta c^{i}_{k-1} \right)^{1 - \gamma_i }}{ 1 - \gamma_i }
\right] ,
\label{compl_opt}
\end{equation}
where the consumption stream $\left( c^{i}_k \right)_{k=0,...,T} $ is a non-negative adapted process satisfying the inequality $c^{i}_k \geq \beta c^{i}_{k-1} ,$ $k=1,...,T,$ and
the budget constraint
\begin{equation}
\sum_{k=0}^{T} E \left[ M_k c^{i}_k \right] =
\sum_{k=0}^{T} E \left[ M_k \e^{i}_k \right] .
\label{budget_constraints}
\end{equation}
Due to the static structure of the habit-forming coefficients, and the fact that the market is complete, the first order conditions (\ref{first_order}) can be re-expressed in a simplified form:
\begin{equation}
\left( c^{i}_{k} - \beta c^{i}_{k-1} \right)^{-\gamma}
= e^{\rho_i} \frac{ \widetilde{M}_k  }{ \widetilde{M}_{k-1}  }
\left( c^{i}_{k-1} - \beta c^{i}_{k-2} \right)^{-\gamma},
\label{simpl}
\end{equation}
for all $k=1,...,T,$ where, in the current setting, the perturbed (aggregate) SPD is given by
\begin{equation}
\widetilde{M}_{k} = M_k +  \sum_{j=1}^{T-k}  \beta^{j}
E \left[ M_{k+j}  \big| \mathcal{G}_{k} \right],
\label{per_spd1}
\end{equation}
for all $k=0,...,T.$ Therefore, the recursive relation
$ \widetilde{M}_{k} = M_k + \beta E \left[ \widetilde{M}_{k+1} \big| \mathcal{G}_k \right],$ $k=0,...,T,$ yields
\[
M_k =  \widetilde{M}_{k} -  \beta E \left[ \widetilde{M}_{k+1} \big| \mathcal{G}_k \right],
\]
for all $k=1,...,T,$ where $\widetilde{M}_{T+1}=0$.

\begin{Definition}
An Arrow-Debreu \textit{equilibrium} (or, \textit{equilibrium} for
short) is a pair of processes $ ( c_{k} )_{ k= 0 , ... , T; i
= 1 , ... , N } ,$ and $ ( M_{k} )_{ k= 1 , ... , T } $ such
that:
\newline
(a) The process $ ( M_{k} )_{ k= 1 , ... , T } $ is a state
price density, and $( c ^{i}_{k} )_{ k= 0 , ... , T } $ is the
optimal consumption stream of each agent $i$ solving the utility
maximization problem (\ref{compl_opt}) in the corresponding market.
\newline
(b) The market clears
 \begin{equation}
\sum_{i=1}^{N} c^{i}_{k} = \epsilon_k :=  \sum_{ i=1}^{N}
\epsilon^{i}_{k},
\label{market_clearing}
\end{equation}
at each period $ k=0 , ... , T$.
\end{Definition}

\subsection{Homogeneous Economy}
We consider now an economy which is populated by an agent of one type,
that is, $N=1$. The risk-aversion, habit-formation parameter, impatience coefficient
and endowments stream are denoted by $\gamma, \beta, \rho$ and $\left( \e_k \right)_{k=0,...,T}$, respectively. 
The associated optimal consumption stream of the individual is denoted by $(c_k)_{k=0,...,T}$.

\begin{Theorem}
In a homogeneous economy, there exists an equilibrium if and only if
\begin{equation}
\e_k > \beta \e_{k-1},
\label{suf1}
\end{equation}
and
\begin{equation}
\left( \e_{k-1} - \beta \e_{k-2} \right)^{-\gamma}
> \beta  e^{ - \rho } E \left[ \left( \e_{k} - \beta \e_{k-1}  \right)^{  - \gamma } \big| \mathcal{G}_{k-1}  \right] ,
\label{suf2}
\end{equation}
for all $k=1,...,T,$ where $\e_{-1}:=0.$
Furthermore, the corresponding equilibrium SPD is unique and given by $M_0=1,$
\begin{equation}
M_k = e^{ - \rho k} \frac{ \left( \e_k - \beta \e_{k-1} \right)^{ - \gamma }
- \beta  e^{ - \rho } E \left[  \left( \e_{k+1} - \beta \e_{k} \right)^{ - \gamma }  \big| \mathcal{G}_k \right]
}{ \e^{-\gamma}_0 - \beta  e^{ - \rho } E \left[  \left( \e_1 - \beta \e_0 \right)^{ - \gamma } \right] } ,
\label{equilib}
\end{equation}
for all $k=1,...,T-1,$ and
\begin{equation}
M_T =
e^{ - \rho T} \frac{ \left( \e_T - \beta \e_{T-1} \right)^{ - \gamma } }{ \e^{-\gamma}_0 - \beta  e^{ - \rho } E \left[  \left( \e_1 - \beta \e_0 \right)^{ - \gamma } \right] } .
\label{equilib2}
\end{equation}
\label{hom_equil}
\end{Theorem}

\begin{Remark}
A sufficient condition for the existence of equilibrium is:
\[
\e_k > \beta \e_{k-1} + \beta^{1 / \gamma } e^{ - \frac{\rho }{ \gamma } } \left( \e_{k-1} - \beta \e_{k-2}   \right) ,
\]
for  all $k=1,...,T.$
\label{remark1}
\end{Remark}

\textbf{Proof of Theorem \ref{hom_equil}.}
We impose the market clearing condition $c_k = \e_k$, for all $k=0,...,T$, which in particular guarantees that budget constraint (\ref{budget_constraints}) is satisfied. By (\ref{simpl}) for $k=1,$ we have
\[
\left( \e_1 - \beta \e_{0} \right)^{ - \gamma } = e^{\rho} \frac{ \widetilde{M}_1 }{ \widetilde{M}_0 } \e^{-\gamma}_0.
\]
Recall that $\widetilde{M}_0 = 1 + \beta E \left[ \widetilde{M}_1 \right] $, hence
\[
E \left[ \widetilde{M}_1 \right]  = \frac{  e^{ - \rho} E \left[ \left( \frac{\e_0}{\e_1 - \beta \e_{0}} \right)^{  \gamma } \right] }{ 1 - \beta e^{ - \rho}
E \left[ \left( \frac{\e_0}{\e_1 - \beta \e_{0}} \right)^{  \gamma } \right] } ,
\]
which yields
\begin{equation}
\widetilde{M}_1 =  \frac{ e^{ - \rho} \left( \e_1 - \beta \e_0  \right)^{-\gamma} }{
\e^{-\gamma}_0 - \beta e^{ - \rho} E \left[ \left( \e_1 - \beta \e_0  \right)^{-\gamma} \right] } .
\label{Mtilde1}
\end{equation}
Next, by (\ref{simpl}) for $k=2$ and the identity $\widetilde{M}_1 = M_1 + \beta E \left[ \widetilde{M}_2 \big| \mathcal{G}_1 \right] $, we get
\[
\left( \frac{ \e_{2} - \beta \e_{1} }{ \e_1 - \beta \e_{0} }  \right)^{-\gamma}  =
e^{ \rho} \frac{ \frac{\widetilde{M}_2 }{ M_{1} } }{ 1 + \beta E \left[ \frac{\widetilde{M}_2 }{ M_{1}}  \big| \mathcal{G}_{k-1} \right] }.
\]
As in (\ref{Mtilde1}), this implies that
\begin{equation}
\frac{\widetilde{M}_2 }{ M_{1}}  =
\frac{ e^{ - \rho} \left( \e_{2} - \beta \e_{1}  \right)^{  - \gamma } }{
\left( \e_{1} - \beta \e_{0} \right)^{-\gamma} - e^{ - \rho} \beta
E \left[ \left( \e_{2} - \beta \e_{1}  \right)^{  - \gamma } \big| \mathcal{G}_{1}  \right] } .
\label{Mtilde2}
\end{equation}
Recall that $\widetilde{M}_1 = M_1 + \beta E \left[ \widetilde{M}_2 \big| \mathcal{G}_1 \right] $.
Therefore, by plugging (\ref{Mtilde2}) into the preceding equation and recalling (\ref{Mtilde1}), we get
\[
M_1 + \beta E \left[  M_1
\frac{ e^{ - \rho} \left( \e_{2} - \beta \e_{1}  \right)^{  - \gamma } }{
\left( \e_{1} - \beta \e_{0} \right)^{-\gamma} - e^{ - \rho} \beta
E \left[ \left( \e_{2} - \beta \e_{1}  \right)^{  - \gamma } \big| \mathcal{G}_{1}  \right] }
\bigg| \mathcal{G}_1 \right] =
\]
\[
 \frac{ e^{ - \rho} \left( \e_1 - \beta \e_0  \right)^{-\gamma} }{
\e^{-\gamma}_0 - \beta e^{ - \rho} E \left[ \left( \e_1 - \beta \e_0  \right)^{-\gamma} \right] },
\]
proving that
\[
M_1 = \frac{ \alpha(1)}{1 + \beta E \left[ \alpha(2) \big| \mathcal{G}_1 \right] } \ \ ; \ \  \widetilde{M}_2 =
\frac{ \alpha(1) \alpha(2) }{
1 + \beta E \left[ \alpha(2) \big| \mathcal{G}_1 \right] } ,
\]
where
\begin{equation}
\alpha(k) := \frac{ e^{ - \rho} \left( \e_{k} - \beta \e_{k-1}  \right)^{  - \gamma } }{
\left( \e_{k-1} - \beta  \e_{k-2} \right)^{-\gamma} - \beta e^{ - \rho}
E \left[ \left( \e_{k} - \beta \e_{k-1}  \right)^{  - \gamma } \big| \mathcal{G}_{k-1}  \right] } ,
\label{alpha}
\end{equation}
for all $k=1,...,T.$
Now, assume that for $k<T-1,$ we have
\begin{equation}
M_k = \frac{ \alpha(1) ... \alpha(k) }{ \prod_{j=2}^{k+1} \left( 1 + \beta E \left[ \alpha(j) \big| \mathcal{G}_{j-1} \right] \right) } .
\label{induction_1}
\end{equation}
Recall that the first order conditions (\ref{simpl}) combined with the identity 
$ \widetilde{M}_k = M_k + \beta E \left[ \widetilde{M}_{k+1}  \big| \mathcal{G}_k \right] $ implies that
\[
\frac{ \frac{\widetilde{M}_{k+1} }{ M_{k}} }{ 1 + \beta E \left[ \frac{\widetilde{M}_{k+1} }{ M_{k}} \big| \mathcal{G}_k \right] } = e^{\rho} \left( \frac{ \e_{k+1} - \beta \e_k }{ \e_k - \beta e_{ k-1} } \right)^{-\gamma} ,
\]
and hence, as in (\ref{Mtilde1}) and (\ref{Mtilde2}), we get
\[
\frac{\widetilde{M}_{k+1} }{ M_{k}}  =
\frac{ e^{-\rho_i} \left( \e_{k+1} - \beta \e_{k}  \right)^{  - \gamma } }{
\left( \e_{k} - \beta  \e_{k-1} \right)^{-\gamma} - \beta e^{-\rho_i}
E \left[ \left( \e_{k+1} - \beta \e_{k}  \right)^{  - \gamma } \big| \mathcal{G}_{k-1}  \right] } ,
\]
thus we get
\begin{equation}
\widetilde{M}_{k+1} = \frac{ \alpha(1) ... \alpha(k+1) }{ \prod_{j=2}^{k+1} \left( 1 + \beta E \left[ \alpha(j) \big| \mathcal{G}_{j-1} \right] \right)}.
\label{induc1}
\end{equation}
As above, we have
\[
\frac{ \frac{\widetilde{M}_{k+2} }{ M_{k+1} } }{ 1 + \beta E \left[ \frac{\widetilde{M}_{k+2} }{ M_{k+1}} \big| \mathcal{G}_{k+1} \right] } = e^{ \rho_i} \left( \frac{ \e_{k+2} - \beta \e_{k+1} }{ \e_{k+1} - \beta e_{ k} } \right)^{-\gamma} ,
\]
and thus
\begin{equation}
\frac{\widetilde{M}_{k+2} }{ M_{k+1}}  =
\frac{ e^{-\rho_i} \left( \e_{k+2} - \beta \e_{k+1}  \right)^{  - \gamma } }{
\left( \e_{k+1} - \beta \e_{k} \right)^{-\gamma} - \beta e^{-\rho_i}
E \left[ \left( \e_{k+2} - \beta \e_{k+1}  \right)^{  - \gamma } \big| \mathcal{G}_{k}  \right] }.
\label{induc2}
\end{equation}
Recall that $\widetilde{M}_{k+1} = M_{k+1} + \beta E \left[ \widetilde{M}_{k+2}  \big| \mathcal{G}_{k+1} \right]$, hence, in virtue of (\ref{induc1}) and (\ref{induc2}), we get
\[
M_{k+1} =
\frac{ \alpha(1) ... \alpha(k+1) }{ \prod_{j=2}^{k+2} \left( 1 + \beta E \left[ \alpha(j) \big| \mathcal{G}_{j-1} \right] \right)},
\]
proving the validity of a similar identity as in (\ref{induction_1}) for $k+1$. The equality $\widetilde{M}_T = M_T$ asserts that
\[
M_T = \frac{ \alpha(1) ... \alpha(T) }{ \prod_{j=2}^{T} \left( 1 + \beta E \left[ \alpha(j) \big| \mathcal{G}_{j-1} \right] \right) } .
\]
Finally, it is not hard to verify that the latter identity and (\ref{induction_1}), $k=1,...,T-1$ yield (\ref{equilib})
and (\ref{equilib2}). The proof is complete. $\qed$

\begin{Remark}
Assume that $\beta=0.$ In this case, the sufficient and necessary conditions
(\ref{suf1}) and (\ref{suf2}) are satisfied, and thus
$M_k = e^{ - \rho k } \left(  \e_k / \e_{0}   \right)^{-\gamma}  $, for all $k=0,...,T$.
\end{Remark}

\subsection{Zero Coupon Bonds and Lucas Tree Equity}
In the current subsection, we compute explicitly the price of a zero coupon bond and the Lucas tree equity, in the setting
of homogenous equilibrium. Moreover, we show that these prices are increasing convex functions of the habit-formation parameter.
For each $T \in N,$ consider a sequence of i.i.d random variables 
$X_1,...,X_T$ such that $X_k > \beta + \beta^{1/\gamma} e^{- \rho / \gamma},$ $P-a.s.$, for each $k=1,...,T.$ 
Assume that the aggregate endowment process $\left( \e_k \right)_{k=0,...,T}$ is a geometric random walk, i.e.,
$\e_o=1$, and 
\[
\e_k = X_1, ... , X_k \ , \ k=1,...,T.
\]
The filtration representing the market is generated by the aggregate endowment process, i.e., $\mathcal{G}_k = \sigma \left( \e_1 , ... , \e_k \right),$ $k=0,...,T.$
Observe that by Remark \ref{remark1}, the sufficient conditions for the existence of an equilibrium 
are satisfied, and thus the SPD is given by (\ref{equilib}) and (\ref{equilib2}). 
\newline
\newline
\textbf{Zero coupon bonds.} 
The price of a \textit{zero coupon bond} at time $k$ maturing at time $m$ is defined by
\[
B^{F}(k,n) = E \left[ \frac{M_n}{M_k} \big| \mathcal{G}_k \right].
\] 
It is not hard to check by using (\ref{equilib}) and (\ref{equilib2}) that
\begin{equation}
B^{F}(k,n) = e^{ - \rho \left( n-k \right) } \frac{ \left( E [ X^{-\gamma}_1 ] \right)^{n-k-1} E [ (X_1 - \beta)^{-\gamma} ]  \left(1-\beta e^{-\rho} E [ X^{-\gamma}_1]  \right)  }{ \left( 1 - \beta / X_k \right)^{ - \gamma} - \beta e^{-\rho}  E \left[ \left( X_1 - \beta \right)^{-\gamma} \right] },
\label{coup1}
\end{equation}
for all $T>n\geq k > 0 ,$
\begin{equation}
B^{F}(0,n) = e^{ - \rho n } \frac{ \left( E [ X^{-\gamma}_1 ] \right)^{n-1} E [ (X_1 - \beta)^{-\gamma} ]  \left(1-\beta e^{-\rho} E [ X^{-\gamma}_1]  \right)  }{ 1 - \beta e^{-\rho}  E \left[ \left( X_1 - \beta \right)^{-\gamma} \right] },
\label{coup2}
\end{equation}
for all $T>n\geq  0 ,$
\begin{equation}
B^{F}(k,T) = e^{ - \rho (T-k) } \frac{ \left( E [ X^{-\gamma}_1 ] \right)^{T-k-1} E [ (X_1 - \beta)^{-\gamma} ]  }{ \left( 1 - \beta / X_k \right)^{-\gamma}
- \beta e^{-\rho}  E \left[ \left( X_1 - \beta \right)^{-\gamma} \right] },
\label{coup3}
\end{equation}
for all $T \geq k \geq 1 ,$ and
\begin{equation}
B^{F}(0,T) = e^{- \rho T } \frac{ \left( E [ X^{-\gamma}_1 ] \right)^{T - 1} E [ (X_1 - \beta)^{-\gamma} ]  }{ 1
- \beta e^{-\rho}  E \left[ \left( X_1 - \beta \right)^{-\gamma} \right] }.
\label{coup3}
\end{equation}
Observe that asymptotically, the yield of the zero coupon bond $ B^{F}(0,T)$ ignores the habit-formation coefficient, namely,
\[
\lim_{T \to \infty } - \frac{ \log B^{F} \left( 0 , T \right) }{ T } =  \rho - \log E \left[ X^{ - \gamma}_1 \right] .
\]
Let us now examine the qualitative behavior of the zero coupon bond $ B^{F}(0,T)$ as a function of $\beta$, for a fixed time horizon.
We fix some $\beta^{*}>0$ and assume that $X_k > \beta^{*} + (\beta^{*})^{1/\gamma} e^{- \rho / \gamma}.$ Note that the zero coupon bond 
$ B^{F}(0,T) ( \beta )$ given by (\ref{coup3}) is well defined for all $\beta \in [0 , \beta^{*}]$. Furthermore, one can check that
\[
\frac{ \partial }{ \partial \beta } B^{F}(0,T) ( \beta ) = 
\]
\[
e^{ - \rho T } \left( E [ X^{-\gamma}_1 ] \right)^{T - 1}
\frac{ e^{-\rho } \left( E \left[ \left( X_1 - \beta \right)^{-\gamma} \right] \right)^2 
+ \gamma E \left[ \left( X_1 - \beta \right)^{-1-\gamma} \right]
}{ \left( 1
- \beta e^{-\rho}  E \left[ \left( X_1 - \beta \right)^{-\gamma} \right] \right)^2 }  > 0,
\]
and
\[
\frac{\partial^2 }{ \partial^2 \beta } B^{F}(0,T) ( \beta ) = 
\frac{ e^{ - \rho T } \left( E [ X^{-\gamma}_1 ] \right)^{T - 1} }{ \left( 1
- \beta e^{-\rho}  E \left[ \left( X_1 - \beta \right)^{-\gamma} \right] \right)^2  } \times
\]
\[
\bigg\{
\left( 1
- \beta e^{-\rho}  E \left[ \left( X_1 - \beta \right)^{-\gamma} \right] \right) 
\bigg( 2  \gamma e^{- \rho} E [ (X_1 - \beta )^{-\gamma} ] E [ (X_1 - \beta )^{-1-\gamma} ]  
+
\]
\[
\gamma \left( 1 + \gamma \right) E [ (X_1 - \beta )^{-2-\gamma} ] \bigg)
+2 \left( e^{-\rho} \left( E [ (X_1 - \beta )^{-\gamma} ] \right)^2  + \gamma E [ (X_1 - \beta )^{-1-\gamma} ] \right) \times
\]
\[
\bigg( e^{-\rho} E [ (X_1 - \beta )^{-\gamma} ] + \beta \gamma e^{-\rho} E [ (X_1 - \beta )^{-1-\gamma} ]  \bigg) 
\bigg\} > 0.
\]
Thus we conclude that the zero coupon bond $ B^{F}(0,T)$ is an \textit{increasing convex} function of $\beta.$

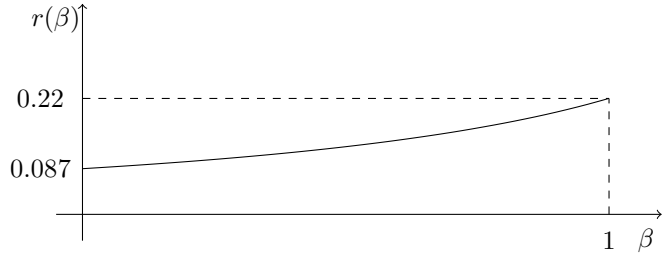
\begin{figure}[h]
\begin{center}
\begin{tikzpicture}[scale=7]

\node (v00x) at (-0.08,0.2203) [circle,draw=blue!0,fill=blue!0,thick,inner sep=0pt,minimum size=2mm] {$0.22$};
\node (v00x) at (-0.08,0.087) [circle,draw=blue!0,fill=blue!0,thick,inner sep=0pt,minimum size=2mm] {$0.087$};
\node (v10x) at (1.07,-0.05) [circle,draw=blue!0,fill=blue!0,thick,inner sep=0pt,minimum size=2mm] {$\beta$};
\node (v10x) at (1,-0.05) [circle,draw=blue!0,fill=blue!0,thick,inner sep=0pt,minimum size=2mm] {$ 1 $};
\node (v10x) at (-0.05,0.37) [circle,draw=blue!0,fill=blue!0,thick,inner sep=0pt,minimum size=2mm] {$ r ( \beta )$};





\draw[draw=black!100,fill=black!0,scale=1,domain=0:1,smooth,variable=\t]
plot ({\t},{ ( (3 - \t)^(-2) + (4 - \t)^(-2) )*( 2 - \t * ( (3 - \t)^(-2) + (4 - \t)^(-2) ))^(-1) });

\draw[dashed] (1,0) -- (1, 0.2203);
\draw[dashed] (0,0.2203) -- (1,0.2203);
\draw[->] (0,-0.05) -- (0,0.4);
\draw[->] (-0.05,0) -- (1.1,0);

\end{tikzpicture}
\caption{Dependence of the interest rate on habit-formation.}\label{fig1:surv}
\end{center}
\end{figure}

\begin{Example}
Consider a market with $T=1$, $\Omega=\{ \omega_1 , \omega_2 \}$, $P( \{ \omega_i \} ) = 1 / 2,$ for $i=1,2$,
$X_1(\omega_1)=3$ and $X_1(\omega_2)=4.$ The agent is represented by $\gamma=2$, $\rho=0$ and the habit-formation
coefficient is some parameter $\beta \in [0,1]$ (that is, $\beta^{*}=1$). As illustrated in Figure \ref{fig1:surv}.,
the zero coupon bond (which is in fact the interest rate) viewed as a function of $\beta$, is given by
\[
r(\beta) := B^{F}(0,1) = \frac{ (3-\beta )^{-2} + (4 - \beta)^{-2} }{ 2 - \beta \left( (3-\beta )^{-2} + (4 - \beta)^{-2} \right) }.
\]

\label{example1}
\end{Example}

${}$
\newline
\textbf{Lucas Tree Equity.} An asset with a dividend being equal to the aggregate endowment is called the Lucas tree equity.
Its price at time $k=0,...,T$ is given by
\[
S^{\e}_{(k,T)} = \sum_{n=k+1}^{T} E \left[ \frac{M_{n}}{M_k} \e_{n} \big| \mathcal{G}_k \right] .
\] 
It is not hard to verify by using (\ref{equilib}) and (\ref{equilib2}) that
\begin{equation}
S^{\e}_{ (k,T)} = \sum_{n=k+1}^{T-1} e^{\rho - ( n - k ) } \frac{
\left( E \left[ X^{1-\gamma}_1 \right] \right)^{n-k-1}
}{ \left( 1 - \beta / X_k \right)^{ - \gamma} - \beta e^{-\rho} E \left[ \left( X_1 - \beta \right)^{-\gamma} \right] }
\times
\label{lucs1}
\end{equation}
\[
\left( E [ X_1 ( X_1 - \beta )^{-\gamma}  ] - \beta e^{-\rho} E [ X^{1-\gamma}_1  ] E [ ( X_1 - \beta )^{-\gamma}  ] \right) \e_k +
\]
\[
e^{ - \rho- ( T - k ) }
\frac{
\left( E \left[ X^{1-\gamma}_1 \right] \right)^{T-k-1} E [ X_1 ( X_1 - \beta )^{-\gamma}  ]
}{ \left( 1 - \beta / X_k \right)^{ - \gamma} - \beta e^{-\rho} E \left[ \left( X_1 - \beta \right)^{-\gamma} \right] } \e_k,
\]
for $k=1,...,T$, and 
\begin{equation}
S^{\e}_{(0,T)}  = 
\frac{1}{  1 - \beta e^{-\rho} E \left[ \left( X_1 - \beta \right)^{-\gamma} \right]}
\bigg( e^{\rho} \frac{1 - \left( e^{-\rho} E \left[ X^{1-\gamma}_1 \right] \right)^{T-1} }{1 - E \left[ X^{1-\gamma}_1 \right] } \times
\label{lucs2}
\end{equation}
\[
\left( E [ X_1 ( X_1 - \beta )^{-\gamma}  ] - \beta e^{-\rho} E [ X^{1-\gamma}_1  ] E [ ( X_1 - \beta )^{-\gamma}  ] \right)+
\]
\[
 e^{-\rho} \left( e^{-\rho} E \left[ X^{1-\gamma}_1 \right] \right)^{T-1} E \left[ X_1 \left( X_1 -\beta \right)^{-\gamma} \right]   \bigg).
\]
Assume that $e^{-\rho} E \left[ X^{1-\gamma}_1 \right] < 1,$ then the long-run Lucas tree equity is given by
\[
S^{\e}_{ (0 , \infty) } := \lim_{T \to \infty } S^{\e}_{ (0 , T) } =  
\frac{ E [ X_1 ( X_1 - \beta )^{-\gamma}  ] - \beta e^{-\rho} E [ X^{1-\gamma}_1  ] E [ ( X_1 - \beta )^{-\gamma}  ] }{ 1 - \beta e^{-\rho} E \left[ \left( X_1 - \beta \right)^{-\gamma} \right]} \frac{e^{\rho}}{ 1 - E \left[ X^{1-\gamma}_1 \right] }.
\]
By a direct computation (similarly as in the case of zero coupon bonds), one can check that $S^{\e}_{( 0 , \infty)} $ is an increasing convex function of the habit-formation parameter $\beta.$

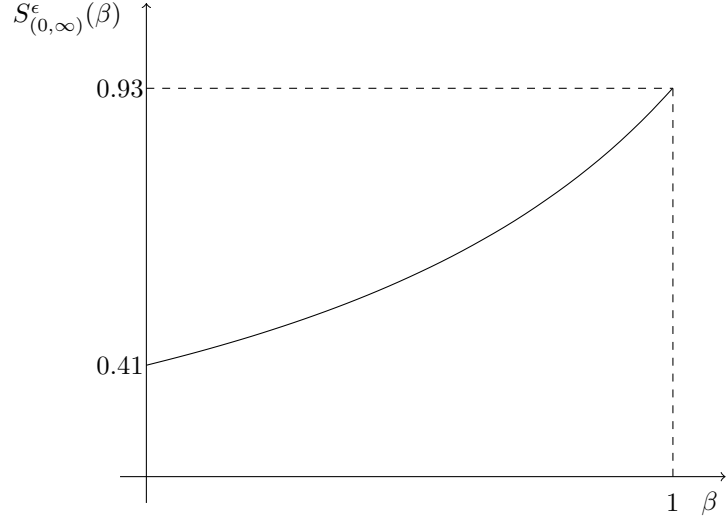
\begin{figure}[h1]
\begin{center}
\begin{tikzpicture}[scale=7]

\node (v00x) at (-0.05,0.4117) [circle,draw=blue!0,fill=blue!0,thick,inner sep=0pt,minimum size=2mm] {$0.41$};
\node (v00x) at (-0.05,0.9381) [circle,draw=blue!0,fill=blue!0,thick,inner sep=0pt,minimum size=2mm] {$0.93$};
\node (v10x) at (1.07,0.15) [circle,draw=blue!0,fill=blue!0,thick,inner sep=0pt,minimum size=2mm] {$\beta$};
\node (v10x) at (-0.15,1.07) [circle,draw=blue!0,fill=blue!0,thick,inner sep=0pt,minimum size=2mm] {$ S^{\e}_{(0 , \infty) } (\beta) $};
\node (v10x) at (1,0.15) [circle,draw=blue!0,fill=blue!0,thick,inner sep=0pt,minimum size=2mm] {$ 1 $};

\draw[draw=black!100,fill=black!0,scale=1,domain=0:1,smooth,variable=\t]
plot ({\t},{ 24/17 * ( ( 3*(3 - \t)^(-2) + 4*(4 - \t)^(-2) )  - 7 /24 * \t * ( (3 - \t)^(-2) + (4 - \t)^(-2) ))*( 2 - \t * ( (3 - \t)^(-2) + (4 - \t)^(-2) ))^(-1) });

\draw[dashed] (1,0.2) -- (1, 0.9381);
\draw[dashed] (0,0.9381) -- (1,0.9381);
\draw[->] (0,0.15) -- (0,1.1);
\draw[->] (-0.05,0.2) -- (1.1,0.2);

\end{tikzpicture}
\begin{center}
\caption{Dependence of the long-run Lucas tree equity on habit-formation.}\label{fig:surv}
\end{center}
\end{center}
\end{figure}

\begin{Example}
Consider the same market as in Example \ref{example1}. Note that in this setting $E[X^{-1}_1] = 7 / 24$. Therefore, the long-run Lucas tree equity is finite
and given by (see Figure \ref{fig:surv}.):
\[
 S^{\e}_{(0 , \infty) } (\beta)= \frac{24}{17} \frac{ \left( 3 - \frac{7}{24} \beta \right)  (3-\beta )^{-2} +
\left( 4 - \frac{7}{24} \beta \right)  ( 4 - \beta )^{-2} 
}{ 2 - \beta \left( (3-\beta )^{-2} + (4 - \beta)^{-2} \right) }.
\]

\label{example2}
\end{Example}

\subsection{Heterogeneous Economy}
By (\ref{simpl}), the optimal consumption stream of agent $i$ satisfies
\begin{equation}
c^{i}_{k} -  \beta c^{i}_{k-1}  =
e^{ - \frac{ \rho_i }{ \gamma_i }  k } \left( \frac{ \widetilde{M}_k }{ \widetilde{M}_{0} } \right)^{ - \frac{1}{ \gamma_i } }
c^{i}_{0} ,
\label{consumption}
\end{equation}
for all $k=1,...,T.$ We emphasize that the system of equations
\begin{equation}
\sum_{i=1}^{N}  \left( c^{i}_k - \beta c^{i}_{k-1} \right) = \e_k - \beta \e_{k-1},
\label{equil_clear}
\end{equation}
for $k=0,...,T$ (where $c_{-1}:= \e_{-1}=0$, $i=1,...,N$), is equivalent to the original
market clearing conditions (\ref{market_clearing}), for $k=0,...,T$.

\begin{Theorem}
Assume that
\begin{equation}
\beta \e_{k} < \e_{k} ,
\label{suf3}
\end{equation}
and
\begin{equation}
 \beta  E \left[  \max_{j=1,...,N} \left( \e_{k} - \beta \e_{k-1} \right)^{-\gamma_j}  \big| \mathcal{G}_{k-1} \right]
<
\min_{j=1,...,N} e^{ - \rho_j } \left( \e_{k-1} - \beta \e_{k-2} \right)^{-\gamma_j} ,
\label{suf4}
\end{equation}
for all $k=1,...,T.$ Then, there exists an equilibrium.
\label{existence_equil}
\end{Theorem}

\textbf{Proof of Theorem \ref{existence_equil}.}
Notice that by (\ref{consumption}) and (\ref{equil_clear}), we can rewrite the budget constraints (\ref{market_clearing}) as
\begin{equation}
\sum_{i=1}^{N} \widetilde{M}^{- \frac{1}{ \gamma_i} }_k e^{- \frac{\rho_i }{\gamma_i} k }
\lambda^{\frac{1}{\gamma_i}}_i =
\e_k - \beta \e_{k-1},
\label{M_k}
\end{equation}
for $k=0,...,T,$ where $ \lambda_i := \left( c^{i}_0 \right)^{\gamma_i} \widetilde{M}_0 $, for $i=1,...,N.$ Recall that $\widetilde{M}_T = M_T $. Since $\e_T > \beta \e_{T-1}$, it follows that exists a continuous function $g_{T} \left( \lambda_1,...,\lambda_N \right) : R^{N}_{++} \to L^{2}_{++} \left( \mathcal{G}_T \right) $ that uniquely solves the equation
\[
\sum_{i=1}^{N} g^{- \frac{1}{ \gamma_i} }_T \left( \lambda_1,...,\lambda_N \right)
e^{- \frac{\rho_i }{\gamma_i} T }
\lambda^{\frac{1}{\gamma_i}}_i =
\e_T - \beta \e_{T-1}.
\]
Thus $M_T = g_T \left( \lambda_1,...,\lambda_N \right) $ is a candidate for a SPD at the maturity date. Recall that $\widetilde{M}_{T-1} = M_{T-1} + \beta E \left[ M_T \big| \mathcal{G}_{T-1} \right] $ and consider equation (\ref{M_k}) for $k=T-1$:
\begin{equation}
\sum_{i=1}^{N} \left( M_{T-1}
+ \beta E \left[ g_T \left( \lambda_1,...,\lambda_N \right) \big|
\mathcal{G}_{T-1} \right]
\right) ^{- \frac{1}{ \gamma_i} }
e^{- \frac{\rho_i }{\gamma_i} \left( T-1 \right) }
\lambda^{\frac{1}{\gamma_i}}_i =
\e_{T-1} - \beta \e_{T-2}.
\label{clear_T-1}
\end{equation}
Note that for each $\omega \in \Omega $ there exists $j(\omega) \in \{ 1 , ... , N \}$ such that
\[
g^{ - \frac{1}{ \gamma_j(\omega)} }_T \left( \lambda_1,...,\lambda_N \right) (\omega)
e^{- \frac{\rho_{j(\omega)} }{\gamma_{j(\omega)}} T }
\lambda^{\frac{1}{\gamma_j(\omega)}}_{j(\omega)} \geq \frac{ \e_T ( \omega) - \beta \e_{T-1} ( \omega)}{N},
\]
or equivalently
\[
 \lambda_{j (\omega) } N^{\gamma_{j ( \omega ) }} e^{- \rho_{j(\omega)} T }
 \left( \e_{T}(\omega) -
 \beta e_{T-1} ( \omega)   \right)^{- \gamma_{j ( \omega)} } \geq g_T \left( \lambda_1,...,\lambda_N \right) (\omega),
\]
which yields
\[
E \left[
\max_{j=1,...,N}
\lambda_j N^{\gamma_j} e^{ - \rho_j T } \left( \e_T - \beta_{T-1} \right)^{ - \gamma_j}
\big|
\mathcal{G}_{T-1}
\right]
\geq
 E \left[ g_T \left( \lambda_1,...,\lambda_N \right) \big|
\mathcal{G}_{T-1} \right] .
\]
Therefore, we have
\[
\sum_{i=1}^{N} \left(
\beta E \left[ g_T \left( \lambda_1,...,\lambda_N \right) \big|
\mathcal{G}_{T-1} \right]
\right) ^{- \frac{1}{ \gamma_i} }
\lambda^{\frac{1}{\gamma_i}}_i e^{- \frac{\rho_i}{\gamma_i} (T-1) } \geq
\]
\[
\sum_{i=1}^{N} \left( \lambda^{-1}_{i} \beta e^{ \rho_i ( T-1 ) } \right)^{ - \frac{1}{ \gamma_i } }
\left(
E \left[
\max_{j=1,...,N}
\lambda_j N^{\gamma_j} e^{- \rho_j T } \left( \e_T - \beta_{T-1} \right)^{ - \gamma_j}
\big|
\mathcal{G}_{T-1}
\right]
\right)^{ - \frac{1}{ \gamma_i } } \geq
\]
\[
\sum_{i=1}^{N}
\left( \lambda^{-1}_{i} \beta e^{ \rho_i ( T-1 ) } \max_{j=1,...,N}  \lambda_j N^{\gamma_j}  e^{ - \rho_j T }  \right)^{ - \frac{1}{ \gamma_i } }
\left(
E \left[
\max_{j=1,...,N}
\left( \e_T - \beta_{T-1} \right)^{ - \gamma_j}
\big|
\mathcal{G}_{T-1}
\right]
\right)^{ - \frac{1}{ \gamma_i } }.
\]
Let $ l \in \{ 1, ... , N \}$ be such that
$ \lambda_l  N^{ \gamma_l } e^{ \rho_l ( T-1 ) } = \max_{j=1,...,N}  \lambda_j N^{\gamma_j}  e^{ \rho_i ( T-1 ) } $. The preceding inequalities yield
\[
\sum_{i=1}^{N} \left(
\beta E \left[ g_T \left( \lambda_1,...,\lambda_N \right) \big|
\mathcal{G}_{T-1} \right]
\right) ^{- \frac{1}{ \gamma_i} }
\lambda^{\frac{1}{\gamma_i}}_i e^{- \frac{\rho_i}{\gamma_i} (T-1) }
>
\]
\[
 \beta^{ - \frac{1}{\gamma_l} }   e^{  \frac{\rho_l}{\gamma_l} }
\left(
E \left[
\max_{j=1,...,N}
\left( \e_T - \beta_{T-1} \right)^{ - \gamma_j}
\big|
\mathcal{G}_{T-1}
\right]
\right)^{ - \frac{1}{ \gamma_l}} >
\e_{T-1} - \beta \e_{T-2},
\]
where the last inequality follows from assumption (\ref{suf4}). Therefore, it follows that
there exists a continuous function $ g_{T-1} \left( \lambda_1 , ... , \lambda_N \right)
: R^{N}_{++} \to L^{2}_{++} \left( \mathcal{G}_{T-1} \right)  $ that solves uniquely the equation:
\[
\sum_{i=1}^{N} \left( g_{T-1} \left( \lambda_1 , ... , \lambda_N \right)
+ \beta E \left[ g_T \left( \lambda_1,...,\lambda_N \right) \big|
\mathcal{G}_{T-1} \right]
\right) ^{- \frac{1}{ \gamma_i} }
e^{- \frac{\rho_i }{\gamma_i} \left( T-1 \right) }
\lambda^{\frac{1}{\gamma_i}}_i =
\]
\[
\e_{T-1} - \beta \e_{T-2}.
\]
Hence, in accordance with (\ref{clear_T-1}), we set
$ M_{T-1} = g_{T-1} \left( \lambda_1 , ... , \lambda_N \right)  $
to be a candidate for a SPD in the period $k=T-1,$ and $\widetilde{M}_{T-1} = \widetilde{g}_{T-1} \left( \lambda_1 , ... , \lambda_N \right) := g_{T-1} \left( \lambda_1 , ... , \lambda_N \right) + E \left[
g_{T} \left( \lambda_1 , ... , \lambda_N \right)  \big| \mathcal{G}_{T-1} \right]  $ is the corresponding perturbed SPD.
Now, one can check by induction that there exists a sequence of continuous functions 
$ g_{0} \left( \lambda_1,...,\lambda_N \right) , ... , g_{T} \left( \lambda_1,...,\lambda_N \right) $ and 
$ \widetilde{g}_{0} \left( \lambda_1,...,\lambda_N \right) , ... , \widetilde{g}_{T} \left( \lambda_1,...,\lambda_N \right) $ such that 
$g_{k} $ $ \big( \lambda_1,...,\lambda_N \big) : R^{N}_{++} \to L^{2}_{++} \left(  \mathcal{G}_k \right) $, and
\[
\widetilde{g}_{k} \left( \lambda_1,...,\lambda_N \right)
=
g_{k} \left( \lambda_1,...,\lambda_N \right) +
\beta E \left[
\widetilde{g}_{k+1} \left( \lambda_1,...,\lambda_N \right)
\big| \mathcal{G}_{k} \right] ,
\]
for all $k=1,...,T-1,$ where $ \widetilde{g}_T \left( \lambda_1,...,\lambda_N \right) :=
g_T \left( \lambda_1,...,\lambda_N \right) ,$ and
\begin{equation}
\sum_{i=1}^{N} \left(
\widetilde{g}_k \left( \lambda_1,...,\lambda_N \right)
 \right)^{-\frac{1}{\gamma_i}}  \lambda^{\frac{1}{\gamma_i}}_i
 e^{ - \frac{ \rho_i }{ \gamma_i } k }
 = \e_{k} - \beta \e_{k-1},
\label{star}
\end{equation}
for all $k=0,...,T.$ A candidate for the (non-normalized) SPD is $M_k = g_{k} \left( \lambda_1,...,\lambda_N \right)$, for $k=0,...,T;$ the corresponding perturbed SPD is
$ \widetilde{M}_k = \widetilde{g}_{k} \left( \lambda_1,...,\lambda_N \right)$, for $k=0,...,T.$
We introduce now the so-called excess demand function $ h := ( h_1 , ... , h_N ) : R^{N}_{++} \to R^{N} ,$ which is defined by
\[
h_{i} ( \lambda_1 , ... , \lambda_N ) =
\frac{1}{ \lambda_i } \bigg(
\sum_{k=0}^{T} E \left[
g_{k} ( \lambda_1 , ... , \lambda_N  )
\sum_{l=0}^{k} \beta^{k-l} e^{ - \frac{ \rho_i }{ \gamma_i } l } \left( \widetilde{g}_l \left(  \lambda_1 , ... , \lambda_N \right)  \right)^{- \frac{1}{ \gamma_i } }
 \right] \lambda^{\frac{1}{\gamma_i}}_i
\]
\[
- \sum_{k=0}^{T} E \left[ g_{k} ( \lambda_1 , ... , \lambda_N )  \e^{i}_k \right]
\bigg),
\]
for $i=1,...,N.$ Observe that by (\ref{consumption}), an appropriate candidate for the optimal consumption stream is $c^{i}_k = \sum_{l=0}^{k} \beta^{k-l} e^{ - \frac{ \rho_i }{ \gamma_i } l } \left( \widetilde{g}_l \left(  \lambda_1 , ... , \lambda_N \right)  \right)^{- \frac{1}{ \gamma_i } } \lambda^{\frac{1}{\gamma_i} }_i$. Furthermore, note that in order to get an equilibrium, it is left to check that the budget constraints (\ref{budget_constraints}) are satisfied for the candidates of the SPD constructed above. Consequently, it suffices to prove that there exists a vector $( \lambda^{*}_1 , ... , \lambda^{*}_N ) \in R^{N}_{++} $ such that $h_{i} ( \lambda^{*}_1 , ... , \lambda^{*}_N ) = 0,$ for all $i=1,...,N.$ To this end, it is sufficient to check that the following properties are satisfied (by a standard fixed-point argument as in Theorem 17.C.1 in Mas-Colell et al. \cite{MWG}):
\newline
\newline
(1) Each function $ h_i,$ $i=1,...,N$ is homogeneous of degree 0
(i.e., $ h_i ( t \lambda_1 , ... ,  t \lambda_N ) = h_i ( \lambda_1 , ... , \lambda_N ) ,$
for all $ ( \lambda_1 , ... , \lambda_N ) \in R^{N}_{++}$ and $t>0$). This follows from
the fact that $g_k,$ $k=0,...,T$ and $\widetilde{g}_{k},$ $k=0,...,T,$ are homogeneous
of degree 1 (i.e., $ g_k ( t \lambda_1 , ... ,  t \lambda_N ) = t g_k ( \lambda_1 , ... , \lambda_N ) ,$ and $  \widetilde{g}_k ( t \lambda_1 , ... ,  t \lambda_N ) = t \widetilde{g}_k ( \lambda_1 , ... , \lambda_N ) ,$ for all $ ( \lambda_1 , ... , \lambda_N ) \in R^{N}_{++}$ and $t>0$).
\newline
\newline
(2) The equation $\sum_{i=1}^{N} \lambda_i h_i ( \lambda_1 , ... , \lambda_N ) = 0 $
holds for all $ ( \lambda_1 , ... , \lambda_N ) \in R^{N}_{++}.$
This is satisfied by the market clearing condition.
\newline
\newline
(3) Each function $h_i,$ $i=1,...,N$ is continuous in $R^{N}_{++}.$ The assertion follows from the fact that
the functions $g_k,$ $k=0,...,T$ and $\widetilde{g}_{k},$ $k=0,...,T$ are continuous.
\newline
\newline
(4) Each function $h_i,$ $i=1,...,N$ is bounded in above and $\lim_{ \lambda_i \to 0}
h_i \left( \lambda_1 , ... , \lambda_N \right) = - \infty$. First, we claim that
\begin{equation}
\lim_{ \lambda_i \to 0 }
 - \frac{1}{ \lambda_i } \sum_{k=0}^{T} E \left[ g_{k} ( \lambda_1 , ... , \lambda_N )  \e^{i}_k \right]  = - \infty.
 \label{limit1}
\end{equation}
Note that by (\ref{star}), we have
\[
\widetilde{g}_{k} ( \lambda_1 , ... , \lambda_N) \geq  \lambda_j e^{ - \rho_j k }
\left( \e_k - \beta \e_{k-1} \right)^{ - \gamma_j } ,
\]
for all $k=0,...,T$ and $j=1,...,N.$ In particular, we have
\[
g_{T} ( \lambda_1 , ... , \lambda_N) =
\widetilde{g}_{T} ( \lambda_1 , ... , \lambda_N) \geq
 \lambda_j e^{ - \rho_j T }
\left( \e_T - \beta \e_{T-1} \right)^{ - \gamma_j },
\]
for all $j=1,...,N$, hence,
\[
g_{T} ( \lambda_1 , ... , \lambda_N)
\geq \frac{1}{N} \sum_{j=1}^{N} \lambda_j  e^{ - \rho_j T }
\left( \e_T - \beta \e_{T-1} \right)^{ - \gamma_j } .
\]
Consider the random variable $ K := \frac{1}{N} \min_{j=1,...,N} \left( \e_T - \beta \e_{T-1} \right)^{ - \gamma_j }
e^{ - \rho_j T }.$ We have
\[
g_{T} ( \lambda_1 , ... , \lambda_N)
\geq K \sum_{j=1}^{N} \lambda_j,
\]
and thus
\begin{equation}
 - \frac{1}{ \lambda_i } \sum_{k=0}^{T} E \left[ g_{k} ( \lambda_1 , ... , \lambda_N )  \e^{i}_k \right]
\leq
- \frac{1}{ \lambda_i }  E \left[ g_{T} ( \lambda_1 , ... , \lambda_N )  \e^{i}_T \right] \leq
\label{inequalitya}
\end{equation}
\[
 -  E \left[ K \e^{i}_T \right] \frac{ \sum_{j=1}^{N} \lambda_j }{\lambda_i} ,
\]
proving (\ref{limit1}).
\newline
\newline
\textit{First case: $0 \leq \gamma_i \leq 1$.} First observe that
\[
\beta^{k-l} E \left[ g_k ( \lambda_1 , ... , \lambda_N )
\big| \mathcal{G}_l \right]
\leq \widetilde{g}_{l} ( \lambda_1 , ... , \lambda_N ) ,
\]
for all $l=0,...,k.$ Thus
\begin{equation}
\sum_{k=0}^{T} E \left[
g_{k} ( \lambda_1 , ... , \lambda_N  )
\sum_{l=0}^{k}  \beta^{k-l} \left( \widetilde{g}_l \left(  \lambda_1 , ... , \lambda_N \right)  \right)^{- \frac{1}{ \gamma_i } }
 \right] \lambda^{\frac{1}{\gamma_i} - 1 }_i \leq
\label{equation0}
\end{equation}
\[
\sum_{k=0}^{T} \sum_{l=0}^{k}
E \left[
\left( \widetilde{g}_l \left(  \lambda_1 , ... , \lambda_N \right)  \right)^{ 1 - \frac{1}{ \gamma_i } }
 \right] \lambda^{\frac{1}{\gamma_i} - 1 }_i.
\]
By (\ref{star}), we have $ \left(  \widetilde{g}_{k} ( \lambda_1 , ... , \lambda_N )  \right)^{ - \frac{1}{ \gamma_i} }  \lambda^{ \frac{1}{ \gamma_i } }_{i} e^{ - \frac{ \rho_i }{ \gamma_i } k }
\leq \e_k - \beta \e_{k-1}$ for all
$k=0,...,T,$ and since $0 \leq \gamma_i \leq 1$, it follows that
\[
\lambda^{ \frac{1}{\gamma_i} - 1}_i \left(  \widetilde{g}_{k} ( \lambda_1 , ... , \lambda_N )  \right)^{1 - \frac{1}{ \gamma_i} }  \leq
e^{ \rho_i k \left( 1 / \gamma_ i -1 \right) }
\left( \e_k - \beta \e_{k-1} \right)^{1 - \gamma_i } ,
\]
and thus
\[
\sum_{k=0}^{T} E \left[
g_{k} ( \lambda_1 , ... , \lambda_N  )
\sum_{l=0}^{k} \beta^{k-l} \left( \widetilde{g}_l \left(  \lambda_1 , ... , \lambda_N \right)  \right)^{- \frac{1}{ \gamma_i } }
 \right] \lambda^{\frac{1}{\gamma_i} - 1 }_i \leq
\]
\[
\sum_{k=0}^{T} \sum_{l=0}^{k}
e^{ \rho_i l \left( 1 / \gamma_ i -1 \right) }
 E \left[ \left( \e_l - \beta \e_{l-1} \right)^{1 - \gamma_i }  \right]  ,
\]
proving that $h_i$ is bounded in above, since the second term is negative.
Furthermore, by (\ref{limit1}) it follows that $\lim_{ \lambda_i \to 0}
h_i \left( \lambda_1 , ... , \lambda_N \right) = - \infty$.
\newline
\newline
\textit{Second case: $ \gamma_i > 1$.} Note that for every $k=0,...,T$ and $\omega \in \Omega,$
there exists an index $j:=j(\omega) \in \{ 1 , ... , N\} $ such that
\[
\left( \widetilde{g}_{k} ( \lambda_1 , ... , \lambda_N ) ( \omega ) \right)^{- \frac{1}{\gamma_j} }
\lambda^{\frac{1}{\gamma_j}}_j >
e^{ \frac{ \rho_i }{ \gamma_i } k }
\frac{ \e_k ( \omega ) - \beta \e_{k-1} ( \omega ) }{N},
\]
or equivalently
\[
 \left( N \left( \e_k ( \omega) - \beta \e_{k-1}(\omega) \right)^{-1}  \right)^{ \gamma_j } e^{ - \rho_j k } \lambda_{j} > \widetilde{g}_{k} ( \lambda_1 , ... , \lambda_N ) ( \omega ) ,
\]
therefore (since $\gamma_i > 1$), we have
\[
 \left( N \left( \e_k ( \omega) - \beta \e_{k-1}(\omega) \right)^{-1}  \right)^{ \gamma_j
 \left( 1 - \frac{1}{\gamma_i} \right) } e^{ - \left( 1 - \frac{1}{\gamma_i} \right) \rho_j k } \lambda^{\left( 1 - \frac{1}{\gamma_i} \right)}_{j} > \left( \widetilde{g}_{k} ( \lambda_1 , ... , \lambda_N ) ( \omega )
 \right)^{1 - \frac{1}{\gamma_i}}.
\]
Set
\[
K' :=
\max_{i,j=1,...,N; k = 0 , ... , T}
\left( N \left( \e_k ( \omega) - \beta \e_{k-1}(\omega) \right)^{-1}  \right)^{ \gamma_j
 \left( 1 - \frac{1}{\gamma_i} \right) } e^{ - \left( 1 - \frac{1}{\gamma_i} \right) \rho_j k }
.
\]
We have
\[
\left( \widetilde{g}_{k} ( \lambda_1 , ... , \lambda_N )  \right)^{ 1 - \frac{1}{ \gamma_i } }
\leq K' \lambda^{1 - \frac{1}{ \gamma_i }}_{j} <
K' \left( \sum_{j=1}^{N} \lambda_j  \right)^{1 - \frac{1}{ \gamma_i }} .
\]
Therefore, as in (\ref{equation0}), we get
\[
h_{i} ( \lambda_1 , ... , \lambda_N ) \leq
\frac{1}{ \lambda_i } \bigg(
\sum_{k=0}^{T} \sum_{l=0}^{k}
E \left[
\left( \widetilde{g}_l \left(  \lambda_1 , ... , \lambda_N \right)  \right)^{ 1 - \frac{1}{ \gamma_i } }  \right]
\lambda^{\frac{1}{\gamma_i}}_i
\]
\[
- \sum_{k=0}^{T} E \left[ g_{k} ( \lambda_1 , ... , \lambda_N )  \e^{i}_k \right]
\bigg)
\]
\[
\leq
\frac{1}{ \lambda_i } \bigg(
\sum_{k=0}^{T} \sum_{l=0}^{k}
E \left[
K' \right]  \left( \sum_{j=1}^{N} \lambda_j  \right)^{1 - \frac{1}{ \gamma_i }}
\lambda^{\frac{1}{\gamma_i}}_i
- \sum_{k=0}^{T} E \left[ g_{k} ( \lambda_1 , ... , \lambda_N )  \e^{i}_k \right]
\bigg)
\]
\[
\leq
\sum_{k=0}^{T} \sum_{l=0}^{k}
E \left[
K' \right] \beta^{k-l} \left( \sum_{j=1}^{N} \frac{\lambda_j}{\lambda_i}  \right)^{1 - \frac{1}{ \gamma_i }}
-  E \left[ K \e^{i}_T \right] \sum_{j=1}^{N} \frac{ \lambda_j }{\lambda_i}
,
\]
where the last inequality follows from (\ref{inequalitya}). Therefore, we get
$ \lim_{ \lambda_i \to 0} h_{i} \left( \lambda_1 , ... , \lambda_N \right) = - \infty. $
From other hand, each function of the form $x \mapsto x^{1 - \frac{1}{\gamma_i}} - x $ is bounded in above on the interval $(0,\infty)$. Hence, the function $h_{i} \left( \lambda_1 , ... , \lambda_N \right)$ is bounded in above, and the proof of Theorem \ref{existence_equil} is accomplished. $\qed$
\newline
\newline
\textbf{Acknowledgments.} I would like to thank my supervisor Semyon Malamud for very helpful discussions and for
important remarks on the first version of the manuscript.
Financial support by the Swiss National Science Foundation via the SNF Grant PDFM2-120424/1 is gratefully acknowledged.

\end{document}